\documentclass{article}

\PassOptionsToPackage{numbers, compress}{natbib}

\usepackage[preprint]{neurips_2026}

\usepackage[utf8]{inputenc} 
\usepackage[T1]{fontenc}    
\usepackage[colorlinks,citecolor=blue]{hyperref}       
\usepackage{url}            
\usepackage{booktabs}       
\usepackage{amsfonts}       
\usepackage{nicefrac}       
\usepackage{microtype}      
\usepackage[dvipsnames]{xcolor}         
\usepackage{amsmath}
\usepackage{amsthm}
\usepackage[linesnumbered,ruled,lined,noend]{algorithm2e}
\usepackage{mathtools}
\usepackage{multirow}
\usepackage{subcaption}
\usepackage{tikz}
\usepackage{listings}
\usepackage{float}

\usetikzlibrary{shapes,arrows}

\theoremstyle{plain}

\theoremstyle{remark}

\title{Domain-Independent Game Abstraction \\ using Word Embedding Techniques}

\author{%
  Juho Kim \\
  Computer Science Department \\
  Carnegie Mellon University \\
  \texttt{juhok@cs.cmu.edu} \\
  \And
  Tuomas Sandholm \\
  Computer Science Department, CMU \\
  Strategic Machine, Inc. \\
  Strategy Robot, Inc. \\
  Optimized Markets, Inc. \\
  \texttt{sandholm@cs.cmu.edu} \\
}

\begin{document}

\maketitle

\begin{abstract}
	Many games of interest in the real world are often intractably large, thereby necessitating the use of game abstraction to shrink them in size, typically by many magnitudes.
	Over the last two decades, there have been significant advances in game abstraction; however, the domain-specific nature (usually poker) of much of the prior work prevents those techniques from being easily generalized to other settings without extensively analyzing the game at hand.
	In this paper, we propose a domain-independent approach to game abstraction, which applies word embedding techniques from the field of natural language processing.
	Treating each action as a word and gameplay data as a corpus, word vectors can be trained to represent each action as a real-valued vector, which can then be clustered to facilitate game abstraction.
	We also explore the use of foundational embedding models and show that action embeddings obtained this way can capture a surprising amount of information about the underlying game.
	Experimental results demonstrate that our proposed game abstraction technique is effective, although it does not outperform specialized algorithms tailored to specific games.
\end{abstract}

\section{Introduction}
\label{sec:introduction}

Many games of interest encountered in the real world are often too large to be solved directly using existing techniques, and even many recreational games, such as two-player no-limit Texas hold'em (containing around $10^{165}$ nodes in the game tree~\cite{ganzfriedandsandholm,johanson}), remain far beyond reach.
This presents the need for techniques to shrink a game's size to be a fraction of the original while retaining much of the strategic considerations present in the original game, that is, game abstraction.
Then, the solutions obtained from the abstractions can be lifted back to yield an approximate solution to the original game.
Game abstraction was crucial for the development of superhuman AI agents for poker~\cite{brownandsandholm2018,brownandsandholm2019}, which represent important milestones for solving large games of imperfect information.

In past decades, there have been significant advances in the study of both the theory and application of game abstraction~\cite{sandholm}.
However, one noticeable trait of much of the prior work~\cite{gilpinandsandholm2006,gilpinetal2007,gilpinandsandholm2008,ganzfriedandsandholm} is that they focus heavily on abstracting the game of poker, a common benchmark for computational game solving.
Unfortunately, abstraction techniques tailored for poker often involve the calculation of poker-specific metrics such as expected hand strength, expected hand strength squared, equities, \textit{etc.}, and while there may be room to extend these to some non-poker settings, this cannot be done without extensive analysis of the game at hand.

That said, there is a class of non-poker-specific game abstraction techniques~\cite{gilpinandsandholm2007b,gilpinandsandholm2007} involving a specialized data structure called signal trees.
However, one needs specialized knowledge about the underlying game to construct signal trees efficiently.
Otherwise, one must enumerate through every terminal node in the game tree, which may not be feasible, depending on the size of the game.

Still, much of the prior work on game abstraction shares a core theme that is not poker-specific: game abstraction is facilitated by merging similar actions with isomorphic subgame trees (\textit{e.g.}, dealing in poker).
In this regard, the problem of game abstraction can be said to reduce to how one clusters actions in relevant game nodes.

\subsection{Our contributions}

To address the generalizability gap in the literature, we propose using \textbf{word embedding techniques}~\cite{mikolovetal,penningtonetal,neelakantanetal} from the field of natural language processing (NLP) for game abstraction.
These techniques, using which one can obtain real-vector representations of words, exhibit strong performance on various NLP tasks.
We apply these techniques by treating actions as words and gameplay data as a corpus, enabling us to represent actions with vectors.
Using their vector representations, actions with isomorphic subgame trees in relevant game nodes can be clustered using the $k$-means algorithm.
Our contribution is empirical in nature and is structured in two parts.

First, in order to \textbf{justify} the use of word embedding techniques for game abstraction, we explore their use for generating action embeddings and demonstrate that these exhibit a surprising amount of interpretability.
For instance, embeddings for strategically similar game actions are located closer to each other in the vector space than those that are not.
We also show that the result remains highly interpretable even when the embeddings are produced using foundational embedding models, which are trained on a more general corpus containing vast amounts of data from multiple domains.
However, the use of foundational embedding models requires us to weaken the claim about our technique's domain-independence, as one must be able to come up with a common-sense textual representation of game actions, which may not always be possible.
We contend that this is much simpler than developing specialized abstraction methodologies for specific games.

Second, we \textbf{quantify} the performance of our proposed technique by calculating exploitabilities of Nash equilibria from abstractions when they are lifted to the original game.
In our experiments, we explore using a word vector technique, namely GloVe, and third-party foundational embedding models from OpenAI and Google to generate action embeddings.
We also compare our methodology with a number of baselines.
The results show that our technique can produce effective abstractions, but it does not outperform specialized algorithms tailored to specific games.
Finally, we conclude by discussing our technique's potential impact and use cases.

\section{Background and related work}

In this section, we review basic game-theoretic concepts and game abstraction.

\subsection{Extensive-form games (EFGs)}

An \textit{extensive-form game (EFG)} has a finite set of players $P$ (containing $n$ non-chance players and chance $p_c$), as well as a finite set of histories $H$.
Each $h \in H$ is a sequence of actions taken by each player, and is associated with an actor $p(h) \in P$ and a set of available actions $A(h)$.
We use $h \cdot a = h'$ to denote that history $h'$ is reached by applying $a \in A(h)$ at $h$.
If $p(h) = p_c$, \textit{i.e.}, $h$ is a chance node, then $f_c(h, a)$ gives a fixed probability distribution over $A(h)$.
Every terminal history $z \in Z \subseteq H$ is associated with a utility $u_i(z)$ for each non-chance player $i \in P \setminus \{p_c\}$.

The presence of private information is represented by partitioning histories into information sets $\mathcal I$.
For each $i \in P \setminus \{p_c\}$, let $\mathcal I_i$ be a subset of $\mathcal I$ belonging to $i$.
Here, $i$ cannot distinguish between any $h, h' \in I \in \mathcal I_i$, so $A(h) = A(h')$, and we denote the common set of available actions by $A(I)$.
Each $i \in P \setminus \{p_c\}$ plays with strategy $\sigma_i$, and, for any $I \in \mathcal I_i$, $\sigma_i(I)$ assigns a probability distribution over $A(I)$.
Let $\sigma = (\sigma_1, \ldots, \sigma_n)$ be a strategy profile.
For each $i \in P$, the expected utility of $i$ given that all players follow $\sigma$ is denoted as $u_i(\sigma)$.

\subsection{Nash equilibrium}

A traditional solution concept in game theory is the \textit{Nash equilibrium}, where no player stands to gain by deviating from their strategy, \textit{i.e.}, given a Nash equilibrium $\sigma^* \in \mathcal X$,
\[
	\forall i \in P, \sigma_i \in \mathcal X_i: u_i(\sigma^*) \ge u_i(\sigma_i, \sigma_{-i}^*),
\]
where $\mathcal X = (\mathcal X_i)_{i = 1}^n$ and $\mathcal X_i$ is the strategy space of $i$.
In this paper, we focus our contribution on two-player zero-sum (2p0s) games, where player utilities at every terminal node sum to zero.
One measure of optimality of a strategy profile (\textit{i.e.}, its distance from a Nash equilibrium) in 2p0s games is the \textit{exploitability}, which is the average of the differences in utility players can achieve by deviating from the original strategy profile, \textit{i.e.}, given a strategy profile $\sigma^* \in \mathcal X$,
\[
	\epsilon = \frac{\max_{\sigma_1 \in \mathcal X_1} u_1(\sigma_1, \sigma_2^*) - \min_{\sigma_2 \in \mathcal X_2} u_1(\sigma_1^*, \sigma_2)}2.
\]
Note that a Nash equilibrium is a strategy profile that achieves zero exploitability.

\subsubsection{Tree-form sequential decision processes (TFSDPs)}

An extensive-form game can be analyzed via the framework of \textit{tree-form sequential decision processes (TFSDPs)}~\cite{farinaetal2019}, where each $i \in P \setminus \{p_c\}$ plays its respective TFSDP.
A player's sequence-form strategy for their TFSDP lies inside a sequence-form polytope, a set of vectors whose elements satisfy a fixed linear flow.
Let $\mathcal X$ and $\mathcal Y$ be the sequence-form polytopes of the row and column players, respectively.
Then, given a sparse utility matrix $A$ and a sequence-form strategy profile $(\vec x, \vec y) \in (\mathcal X, \mathcal Y)$, the expected utilities of row and column players can simply be expressed as $u = (\vec x)^\top A\vec y$ and $v = -(\vec x)^\top A\vec y$, respectively.
The utility matrix is highly sparse and can be implemented as such.
The time complexities of solving 2p0s games via linear programming or a single iteration of common 2p0s-game-solving algorithms (\textit{e.g.}, CFR~\cite{zinkevichetal2007}) can be expressed in terms of the number of sequences (\textit{i.e.}, $\dim(\mathcal X) + \dim(\mathcal Y)$) and non-zeros in the utility matrix.
This is important, as we later use these as measures of the game size.

\subsection{Word embedding techniques}

Word embedding techniques have now become ubiquitous in NLP.
They exhibit strong performance for tasks that involve quantifying the semantics of words or texts.
Some of the earlier works in this field include word vector techniques, such as word2vec~\cite{mikolovetal} and GloVe~\cite{penningtonetal}, which, given a large corpus, fit a set of static vectors for every word that appears a sufficient number of times.
These vectors can be thought of as quantifying the underlying words' meaning.
Empirically, the Euclidean or cosine distance between word vectors reflects how similar the corresponding words are in meaning.
Word vectors also exhibit a locally linear substructure.
For example, the difference between the vectors for `man' and `woman' is roughly equivalent to that for `king' and `queen'.

While revolutionary, as of the writing of this paper, word embedding models, based on the design of \textit{large-language models (LLMs)}, have largely superseded word vector techniques in most usage.
These models, incorporating the transformer architecture, leverage attention and can even encode general text and words that are not present in the original vocabulary.
Foundational embedding models are embedding models that are pre-trained on a large set of data across multiple domains.~\cite{neelakantanetal}
In this paper, we explore the use of word vector/embedding techniques for game abstraction.
For simplicity, we call both families of techniques `word embedding techniques'.

\subsection{Game abstraction}

While there have been many advances in both the theory and application of game abstraction~\cite{sandholm}, we focus on the application side of the prior work in this overview.

Abstraction in poker started with Rhode Island hold'em, a small artificial poker variant~\cite{shiandlittman,shiandlittman2001}, which was followed by manual abstraction for simplifying two-player limit Texas hold'em~\cite{billingsetal}.
This, in turn, was followed by the introduction of \textit{GameShrink}~\cite{gilpinandsandholm2007b}, a lossless abstraction algorithm for imperfect-information extensive-form games.
The core idea of GameShrink is to filter the information provided to the players in a lossless way.
This technique is highly suited for poker, as it is usually suit-agnostic in that there is nothing special about a particular suit (\textit{e.g.}, a spade) compared to another.
The authors proposed building a signal tree, where each edge corresponds to a revelation of some information to a player.
In poker, the signals correspond to the dealt cards, and building a signal tree does not entail fully processing the game tree, as betting dynamics between players can be ignored entirely (although recognizing this requires knowledge about the topology of poker game trees).
By testing for the utility equivalence between subgame trees, that is, whether different signals correspond to the same set of utilities in the analogous leaves, one can merge signals without increasing exploitability.

Lossless abstraction, although useful, is often not powerful enough to yield a game whose size is small enough to be solved.
To address this, a follow-up work~\cite{gilpinandsandholm2007} extended GameShrink~\cite{gilpinandsandholm2007b} to merge `similar' but not necessarily equivalent signals: lossy abstraction.
Here, the authors proposed a matching heuristic of using utility differences as a distance metric between different signals.

There still remained a need for the means to find how to optimally merge across different levels of the game tree given a budget (\textit{e.g.}, a fixed number of resulting clusters).
For this,~\citet{gilpinandsandholm2007} proposed integer programming to obtain optimal clusters at each game tree, depth-by-depth.

The next idea was \textit{potential-aware} abstraction~\citep{gilpinetal2007,gilpinandsandholm2008}, which is to bucket the states at each level---starting from the leaves where they are bucketed simply based on hand strength---according to their transition probability vectors to next-level buckets.
This captures a richer notion of hand quality and how it evolves during the play of a poker hand than simple hand strength.

As of the writing of this paper, the state-of-the-art technique for this line of research is potential-aware abstraction, with earth-mover's distance as the distance metric between transition probability vectors~\cite{ganzfriedandsandholm,brownetal}. These were the information-abstraction algorithms used in the superhuman poker AIs \textit{Libratus}~\cite{brownandsandholm2018} and \textit{Pluribus}~\cite{brownandsandholm2019} for two-player and multi-player no-limit Texas hold'em, respectively.

Finally, there are other game abstraction techniques specialized for poker~\cite{gilpinetal2007,ganzfriedandsandholm2008,gilpinetal2008}.
While there are a few general-purpose abstraction techniques that are not poker specific~\cite{sandholmandsingh,kroerandsandholm2014,kroerandsandholm2015,kroerandsandholm2016,kroerandsandholm2018}, they are not very scalable.

\subsection{Shortcomings in the literature}

From the perspective of practical game abstraction with general applicability, there are significant gaps in the literature.
Consider the family of abstraction techniques using signal trees.
Without any specialized knowledge about the underlying game, checking for the utility equivalence between subgame trees requires the full enumeration of their terminal nodes, which is typically impractical.
The same holds for calculating the utility differences between subtrees for the distance metric in lossy abstraction~\cite{gilpinandsandholm2007}.
Additionally, the earth mover's distance cannot be calculated efficiently in high-dimensional settings~\cite{ganzfriedandsandholm}.
Also, the view of poker presented by these abstraction algorithms is arguably \textbf{incomplete} in that only considering card rollouts cannot fully account for the dynamics of how different poker hands interact.
Indeed, poker hands can affect how players act during the betting stages, and some hands can be much more playable in certain betting situations than others.

Our proposal of using word embedding techniques for game abstraction addresses the above gaps.
We can assign a unique token to each action and use the tokens to record gameplay data (simply, a sequence of actions).
The gameplay data can then be processed by word vector techniques to fit the action embeddings.
When applied to poker, since the tokens in the corpus are not exclusively for signals, the action embeddings have the potential to capture betting dynamics as well.
Note that our technique can be applied in a completely domain-independent fashion, and one can even use existing (possibly synthetic) gameplay data.
On top of this, if foundational embedding models are used to generate action embeddings, no training is needed, and embeddings corresponding to different actions (or signals) can be obtained with relative ease.
However, the use of foundational models weakens our claim of domain independence, as one needs to come up with a common-sense method of representing each action using a human-readable text, which may not always be possible.

\section{Obtaining action embeddings using word embedding techniques}
\label{sec:observational-studies}

We are now ready to present our results.
We begin by justifying the use of word embedding techniques for obtaining action embeddings.
For this, we conduct several observational studies, showing that action embeddings obtained in this way can capture deep strategic consideration of the game actions.

\subsection{Observational studies on the $k$-nearest neighbors}

Recall that the Euclidean or cosine distance between a pair of word embeddings gives a good estimate of the semantic similarity of the underlying words.
In this section, we show that an analogue of this result holds even for game playing, where the distance between a pair of action embeddings gives an interpretable representation of their strategic similarity of the underlying actions.
We do so by choosing several target actions and querying their nearest neighbors in the embedding space.

\subsubsection{Chess embeddings}

\begin{figure}[t!]
	\centering
	\includegraphics[width=0.32\linewidth]{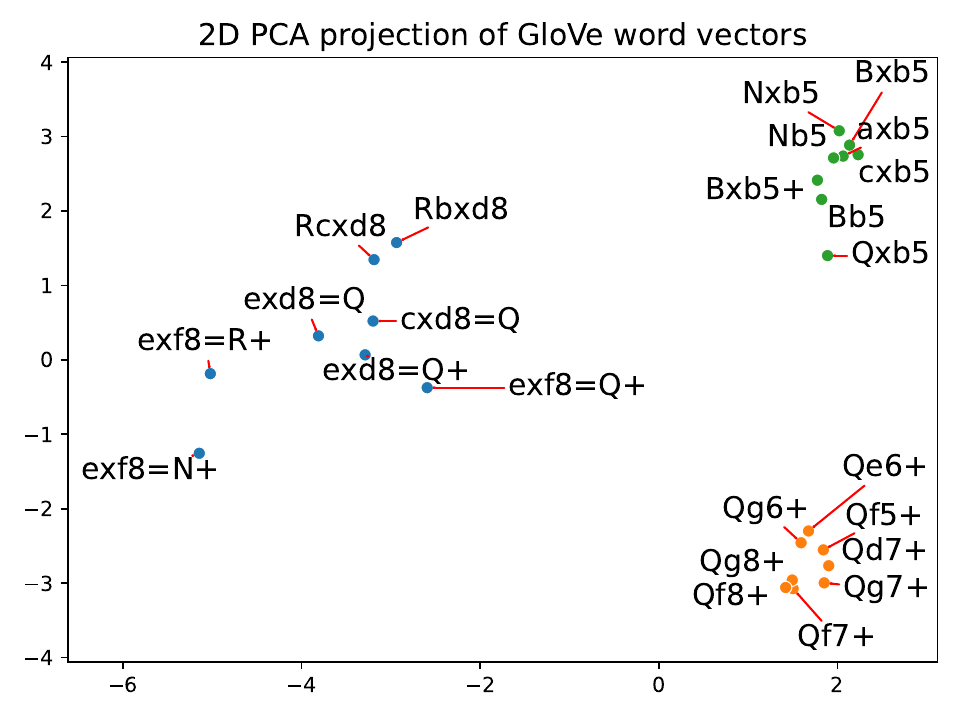}
	\includegraphics[width=0.32\linewidth]{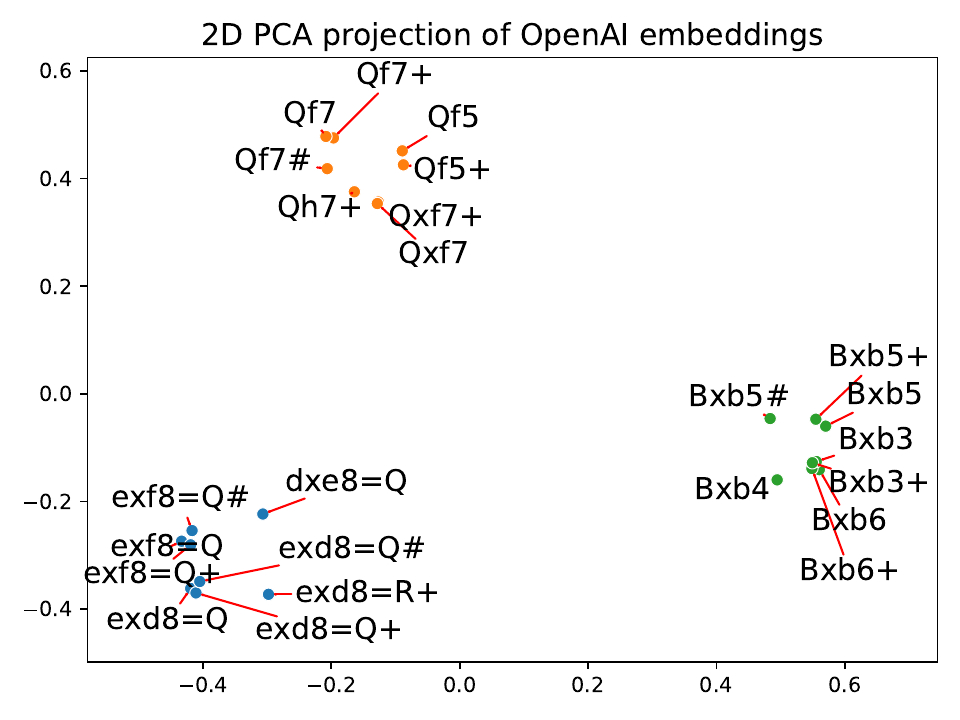}
	\includegraphics[width=0.32\linewidth]{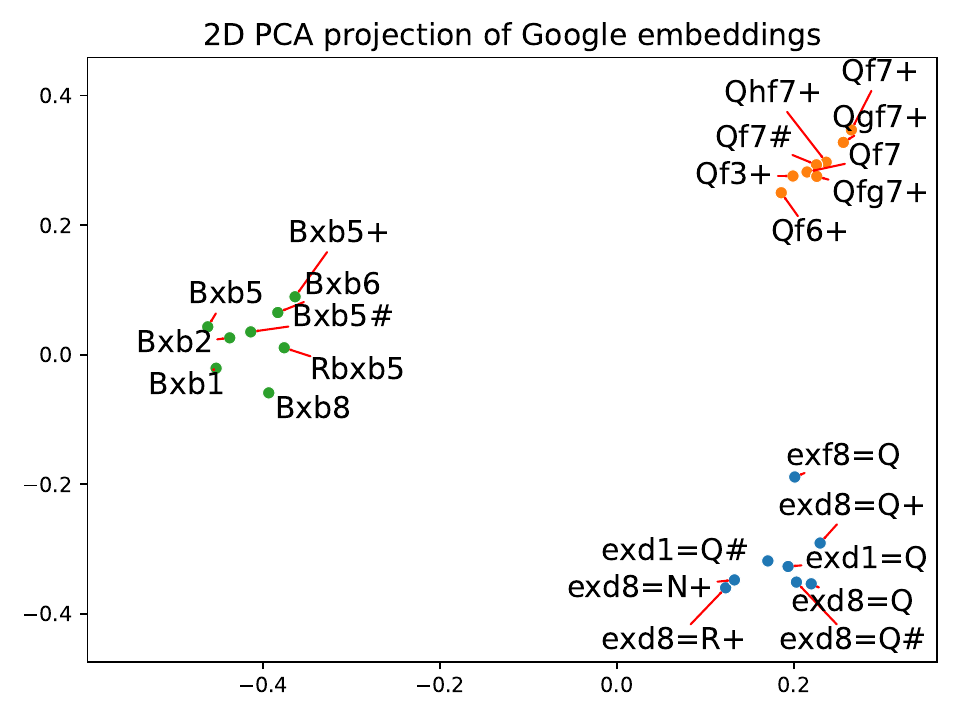}
	\caption{
		Nearest neighbors of actions `\texttt{exd8=Q}', `\texttt{Qf7+}', and `\texttt{Bxb5}' projected using 2D PCA.
		The left, middle, and right figures, respectively, visualize action embeddings obtained from GloVe, OpenAI `\texttt{text-embedding-3-small}', and Google Gemini `\texttt{gemini-embedding-001}', respectively.
		Indeed, chess actions with similar properties are located closer to each other than those that are not.
	}
	\label{fig:knn}
\end{figure}

First, we trained GloVe word vectors on a dataset of official chess games from human experts~\cite{pgnmentor}.
The parameters used during training are given in Appendix~\ref{sec:parameters}.
We used a separate token for every distinct chess action.
Additionally, on the same vocabulary, we obtained embeddings for the chess actions using third-party foundational embedding models from OpenAI and Google Gemini.
The foundational embedding models we used are detailed in Appendix~\ref{sec:foundational-embedding-models}.
For the study, we chose the following target chess actions: `\texttt{exd8=Q}', `\texttt{Qf7+}', and `\texttt{Bxb5}'.
For each action and embedding model, we obtained nearest neighbors, as tabulated in Appendix~\ref{sec:knn}.
We also projected them into 2-dimensional space using \textit{principal component analysis (PCA)}~\cite{jolliffe}, as shown in Figure~\ref{fig:knn}.

We begin with our observation on the results from the action embeddings obtained via GloVe.
For `\texttt{exd8=Q}' (\textit{i.e.}, pawn captures a piece in the d8 square and promotes to queen), we see that 4 of the 7 nearest neighbors also involve capturing with a pawn and promoting to some piece.
Additionally, all of the moves correspond to moving some piece to the top file, and 5 of the 7 actions involve promotion.
More strikingly, all seven nearest neighbors involve capturing a piece.
This highly interpretable result is especially striking because GloVe, as we trained it, is \textbf{unaware} of the individual characters in the action notation that serve as descriptors for a particular action, such as checkmate (denoted by \#), check (+), capture (x), promotion (=Q, =R, \textit{etc.}), moved piece, and coordinates.
The neighbors of other target actions are just as interpretable.
All seven nearest neighbors of `\texttt{Qf7+}' (\textit{i.e.}, queen to f7 check) involve moving a queen to squares near f7 without capturing a piece while checking.
Similarly, nearest actions of `\texttt{Bxb5}' (bishop captures a piece in b5) involve moving some piece (mostly to capture) to b5.
Furthermore, the nearest neighbors are clustered together in the projection plot.

We also obtained highly interpretable results using foundational embedding models.
As shown in the plots in the middle and on the right, target actions are located near other actions with similar or matching qualitative attributes.
Again, the nearest neighbors are tabulated in Appendix~\ref{sec:knn}.
It is worth mentioning that, unlike GloVe, tokenization used by foundational embedding models, based on the design of LLMs, can infer the presence of different attributes (checkmate \#, check, +, \textit{etc.}), so the fact that the results from foundational embedding models are highly interpretable is not as surprising as the findings for GloVe.
That said, this consideration can only help the action embeddings be more useful during game abstraction.

\subsubsection{Poker embeddings}
\label{sec:poker-embeddings}

\begin{figure}[t!]
	\centering
	\begin{subfigure}[t]{0.324\textwidth}
		\centering
		\includegraphics[width=\linewidth]{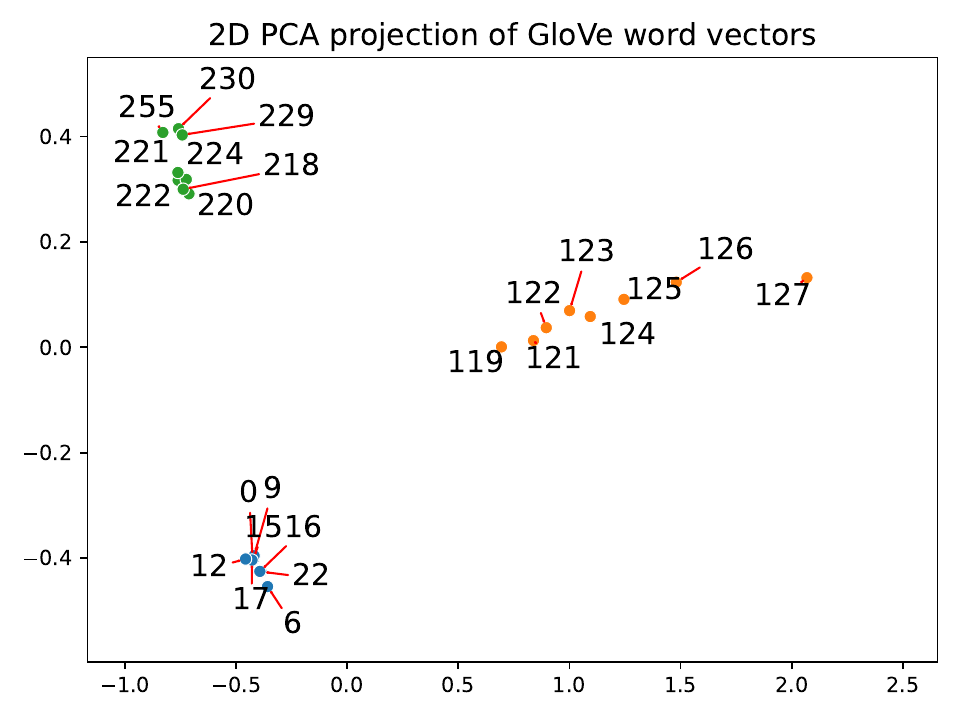}
		\caption{
			Nearest neighbors of dealing the row player the 0\textsuperscript{th}, 127\textsuperscript{th}, and 255\textsuperscript{th} cards in 256-Kuhn poker.
		}
		\label{fig:knn-kuhn}
	\end{subfigure}
	\begin{subfigure}[t]{0.66\textwidth}
		\centering
		\includegraphics[width=0.49\linewidth]{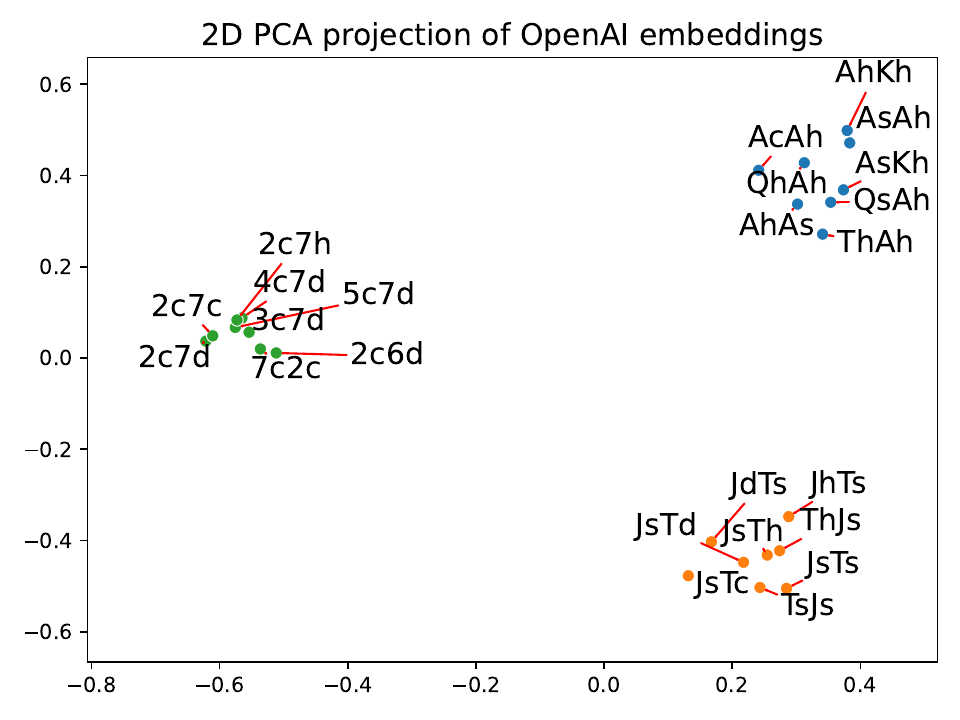}
		\includegraphics[width=0.49\linewidth]{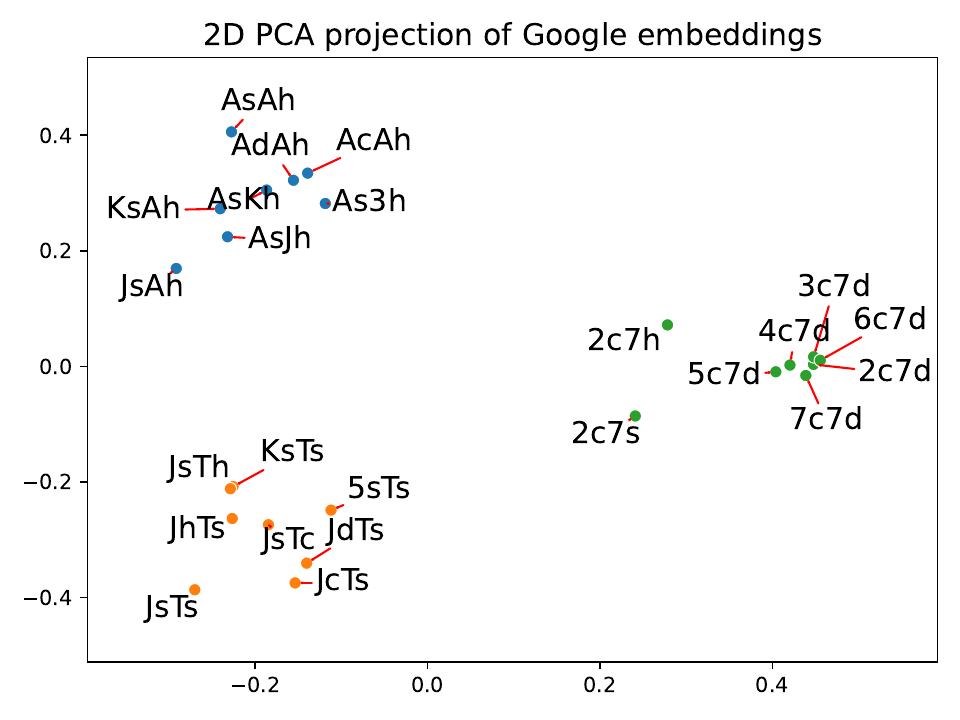}
		\captionsetup{margin={1em, 1em}}
		\caption{
			Nearest neighbors of different pairs of cards in poker: `AsAh', `JsTs', and `2c7d'.
		}
		\label{fig:knn-leduc}
	\end{subfigure}
\end{figure}

Next, we tested whether the findings also apply to poker. In this observation study, we focus on a variant of Kuhn poker~\cite{kuhn}, a common benchmark game, but played with a deck of 256 unique cards instead of 3, henceforth referred to as 256-Kuhn poker; the rules are described in Appendix~\ref{sec:games}.
Unlike normal poker, which consists of 13 ranks (deuces to aces), there is no accepted way to represent this many ranks using text.
Thus, we do not use foundational embedding models for this particular study, but instead only use GloVe to do so, with each token representing a distinct action (or chance events).
We solved for a Nash equilibrium of 256-Kuhn poker and generated a billion play-throughs using the calculated strategy profile. Then we used that data to fit the GloVe word vectors.
The training parameters are detailed in Appendix~\ref{sec:parameters}.

As we did for chess above, we selected several target actions, specifically, dealing the row player the 0\textsuperscript{th}, 127\textsuperscript{th}, and 255\textsuperscript{th} cards, and obtained the nearest neighbors (tabulated in Appendix~\ref{sec:knn-kuhn}).
Their 2D PCA projections are shown in Figure~\ref{fig:knn-kuhn}.
One can observe that the procedure does a good job of clustering the hole cards with similar ordinals.
Most notably, the corresponding ordinals of the nearest neighbors of dealing the 127\textsuperscript{th} card to the row player range from 119 to 126, and thus are tightly distributed.
Grouping poker hands of similar strength is generally how state-of-the-art abstraction techniques specialized for poker operate, and our observational study suggests that obtaining action embeddings via word embedding techniques has the potential to be effective for poker as well.

Next, we explore the use of foundational embedding models for Texas hold'em, a variant of poker.
Unlike chess, there is no widely accepted action notation for betting actions in poker, such as fold, check, call, \textit{etc}.
However, there is a standardized way to denote a set of cards in poker (see Appendix~\ref{sec:poker-hands}), which can nonetheless be used for card dealings.
Using the aforementioned foundational models from OpenAI and Google Gemini, we fetched the embeddings of the textual representation of different poker hands.
We then selected several target poker hands---`AsAh', `JsTs', and `2c7d'---and obtained their nearest neighbors (tabulated in Appendix~\ref{sec:knn-leduc}).
Their 2D PCA projections are shown in Figure~\ref{fig:knn-leduc}.
The pair of Aces we chose as a target is located close to other pairs of Aces or hands that include an Ace, which are strong hands in the Texas hold'em variant of poker.
It can also be seen that `JsTs' is grouped with other hands containing a jack and a ten, which are hands of similar strength in Texas hold'em.
Also, it is notable that `2c7d' was grouped with other borderline hands.

While our results using foundational embedding models show that they can group poker hands of similar strength together, we noticed that foundational embedding models also encode information about the suits, which is pertinent in some situations in poker but irrelevant in others.
Indeed, with specialized knowledge about the game being played, it is likely that one could come up with a rule-based system for generating abstraction embeddings that are better than what can be obtained using foundational embedding models.
That said, our focus is on developing domain-independent game abstraction techniques, and our results suggest that generalized action embeddings from foundational embedding models can yield effective game abstractions.

\section{Experiments on play quality}

In our experiments on play quality, we generated game abstractions using action embeddings obtained from word embedding techniques.
We use the following two benchmark games: 256-Kuhn Poker and 13-Leduc hold'em (defined in Appendix~\ref{sec:games}) to demonstrate the practicality of our technique.

We measure the quality of the game abstraction primarily in two ways.
The first measures the degree of optimality of a Nash equilibrium from the abstraction when it is \textbf{lifted} to the original game.
In game theory, a standard metric for the quality of a strategy profile in imperfect-information games is exploitability, which we calculate and plot in our experiments.
The second measure is the size of the resulting abstraction.
While there are different ways to measure the game size, we use the number of sequences and the number of non-zeros (NNZs) in the utility matrix to quantify the game size, as the time complexities of common methods for solving imperfect-information games, such as linear programming and sequence-form counterfactual regret minimization, can be expressed with respect to these values.
Note also that the NNZs in the utility matrix are asymptotically equal to the number of terminal nodes.
To summarize, the two measures of importance for abstraction are exploitability and the size of the resulting abstract game, both of which we prefer to be small.

Throughout our experiments, we used the following two baselines to compare against: the hand-bucketing algorithm and the random baseline.
The hand-bucketing algorithm is a widely established way to abstract poker games.
It a) accepts the number of buckets as a parameter, b) sorts the hole cards (\textit{i.e.}, observations) in the order of their strength, and c) partitions the ordering into uniformly-sized buckets.
The random baseline accepts the number of clusters and a random seed as parameters, and assigns actions to clusters uniformly at random.

To demonstrate that our abstraction technique results in an effective abstraction, we expect our technique to outperform the random baseline.
That said, we do not expect our much more general technique to outperform the hand-bucketing baseline, tailored specifically for poker.
We also expect the game abstractions to typically yield better lifted strategies, the finer-grained the abstractions are (although game abstraction pathologies do sometimes exist: having a finer-grained abstraction can sometimes hurt exploitability~\cite{waughetal}).
Since the hand-bucketing algorithm can only cluster actions in chance nodes, we also restrict our proposed methodology and the random baseline to cluster actions only in these nodes in order to make a fair comparison.

\subsection{Using GloVe word vectors}

\begin{figure}[t!]
	\centering
	\includegraphics[width=0.4\linewidth]{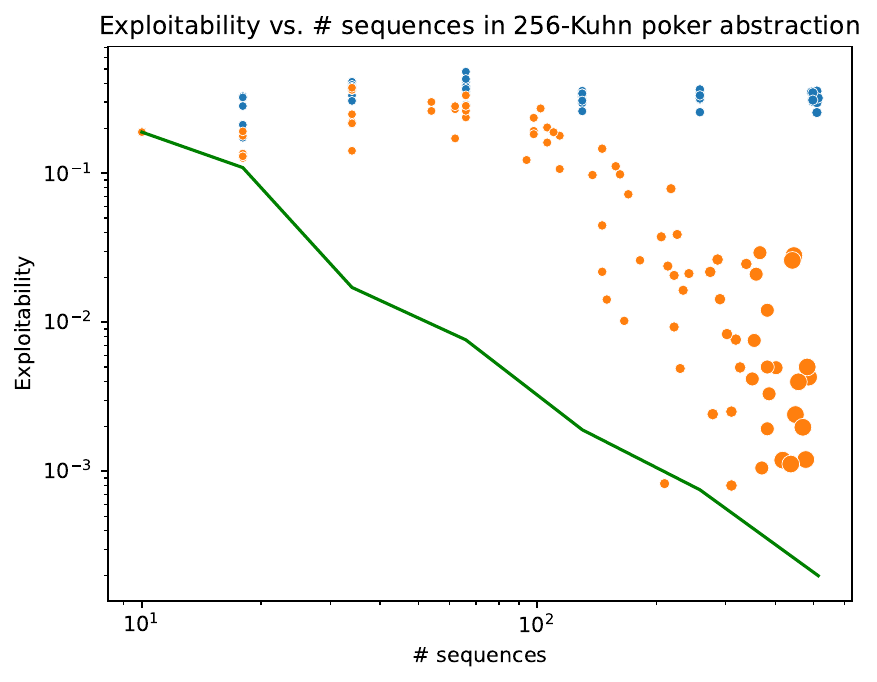}
	\includegraphics[width=0.51\linewidth]{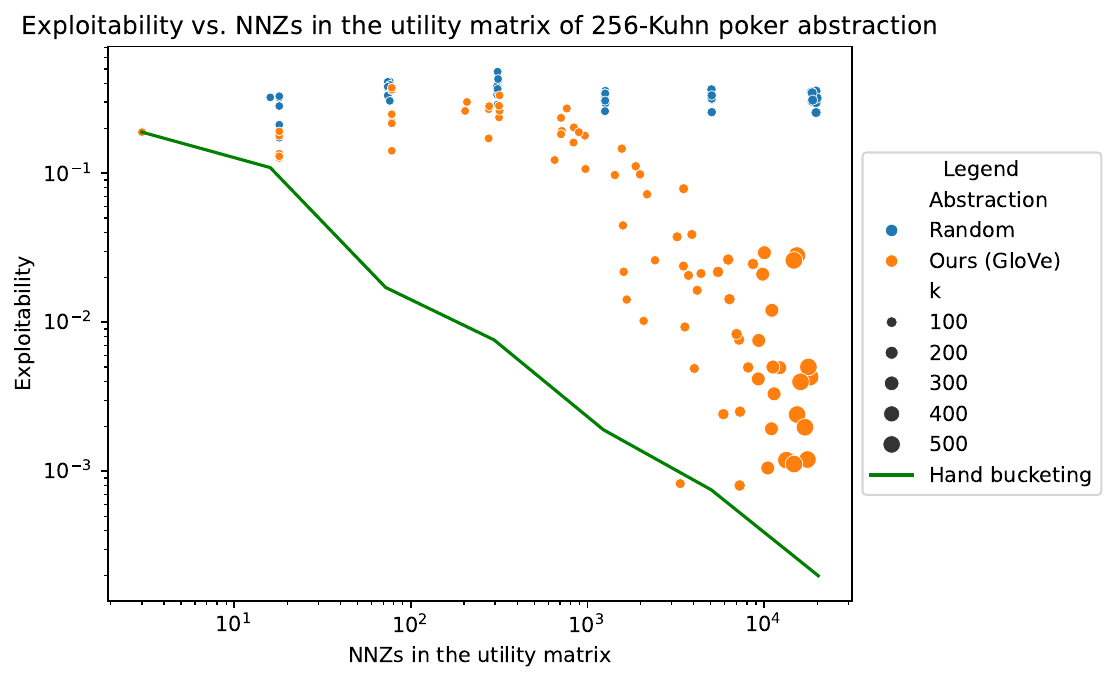}
	\caption{
		Exploitability versus abstraction size for 256-Kuhn poker.
		Each dot represents one generated abstraction.
		The two plots show the same set of results, but the left one uses the number of sequences for the abstraction size, whereas the right one uses the NNZs in the utility matrix.
	}
	\label{fig:exploitability}
\end{figure}

We begin by demonstrating the practicality of using GloVe word vectors to generate action embeddings using 256-Kuhn poker.
We used the chess action embeddings we generated using GloVe for 256-Kuhn poker from our observational study in Section~\ref{sec:poker-embeddings}.
256-Kuhn poker has 512 chance events, one for each player and possible hole card to be dealt.
To generate our abstractions, we ran $k$-means clustering with exponentially increasing values of $k$, from $1$ to $512$.
Also, for each choice of $k$, we generated ten separate abstractions with varying random seeds.
Similarly, for the random baseline, we varied the number of clusters using the same $k$ values and random seeds.
The same $k$ values were also used as the number of buckets for the hand-bucketing baseline.

The exploitability plots for the different abstraction methodologies on 256-Kuhn poker are shown in Figure~\ref{fig:exploitability}, which shows how exploitability varies as the size of the game changes.
We produced two plots on the same data, as we used two different metrics for the size of the abstraction.
As expected, for our proposed methodology and the hand-bucketing baseline, an increase in the size of the abstraction generally corresponds to a decrease in exploitability.
Intuitively, having finer-grained abstraction is beneficial in that it improves the quality of the lifted solution.
However, increasing the size of the abstraction did not help with the performance of the random baseline.

Another observation to note is that the performance of our technique generally outperforms the random baseline, and the difference widens as the resulting abstraction size increases.
We use this as empirical evidence that generating action embeddings for game abstraction using word vector techniques such as GloVe can yield reasonably good game abstractions.
That said, we also see that the hand-bucketing baseline, tailored for poker, outperforms our technique; this is expected since our technique is domain-independent and does not rely on specialized knowledge about the game.

\subsection{Using foundational embedding models}

\begin{figure}[t!]
	\centering
	\includegraphics[width=0.4\linewidth]{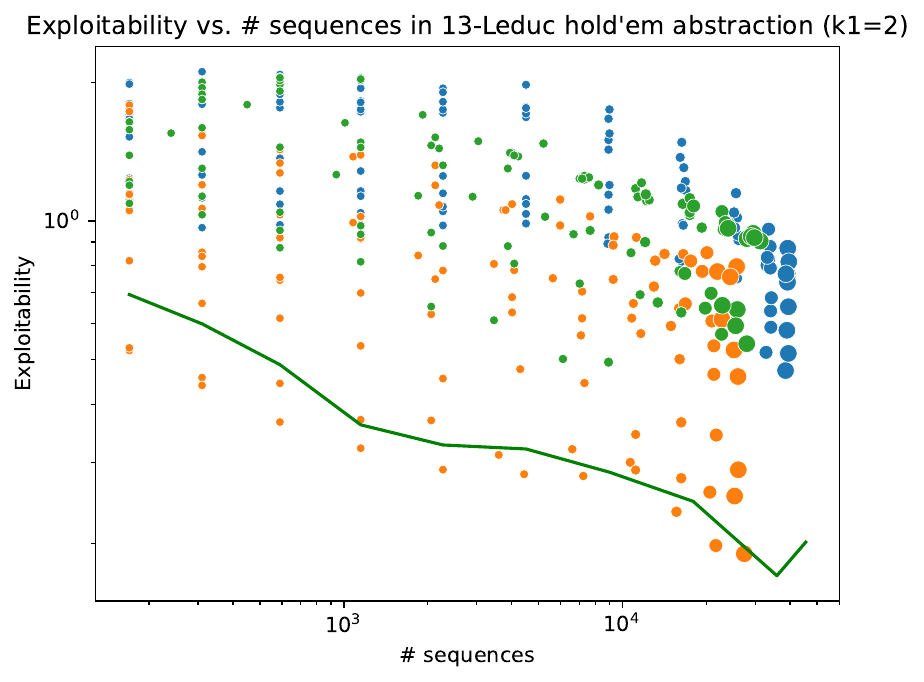}
	\includegraphics[width=0.496\linewidth]{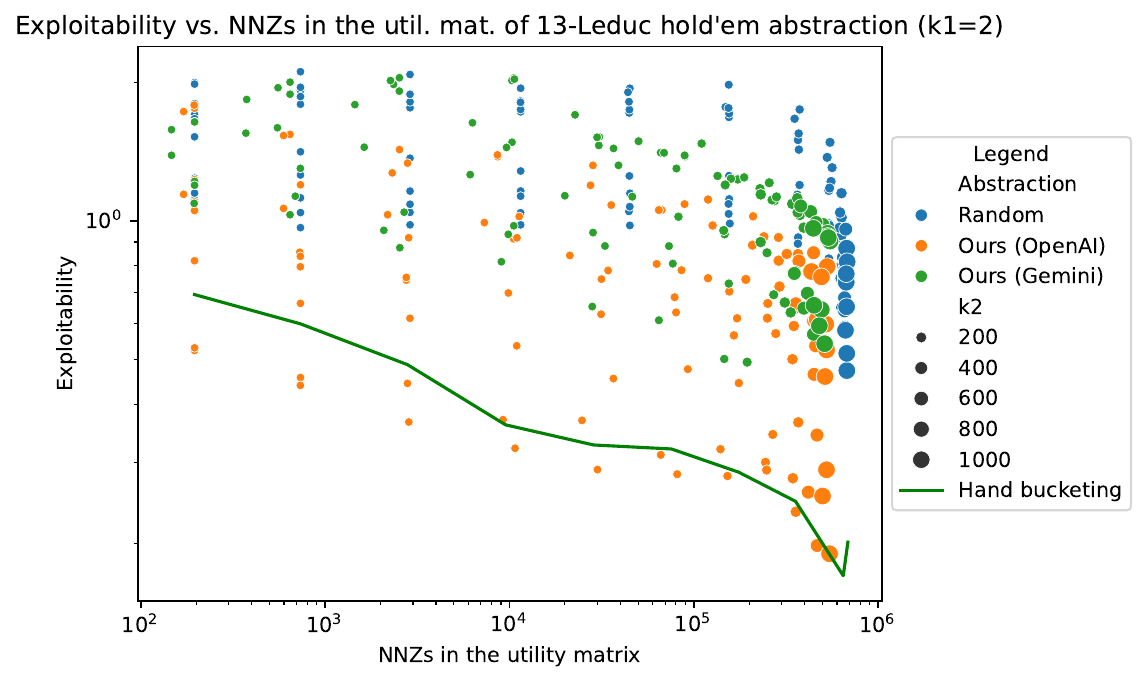}
	\includegraphics[width=0.4\linewidth]{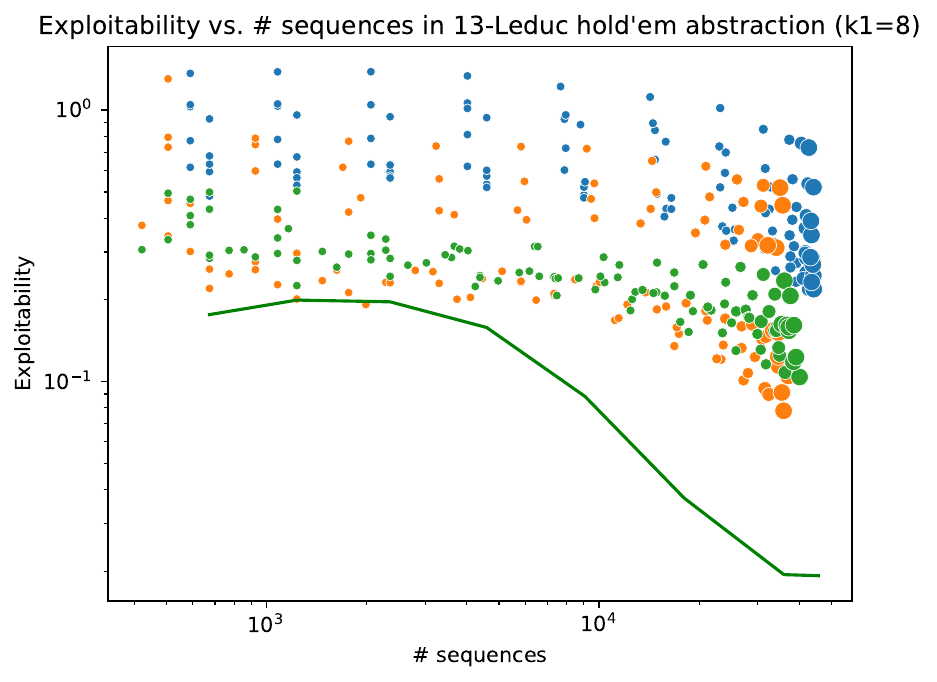}
	\includegraphics[width=0.49\linewidth]{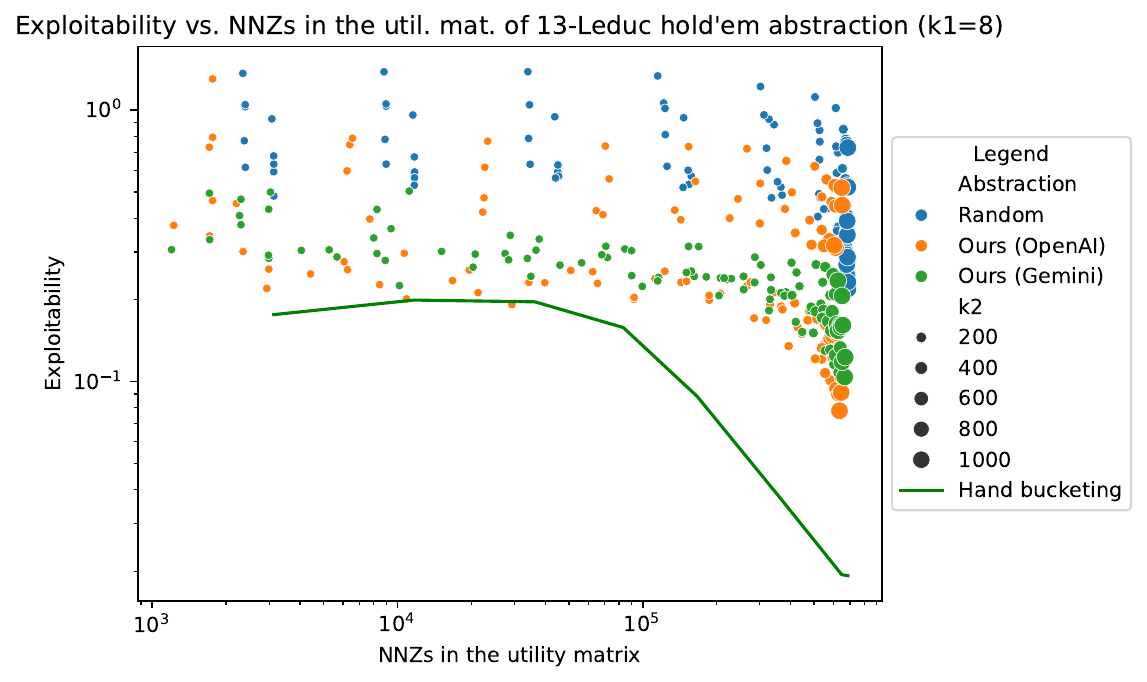}
	\includegraphics[width=0.4\linewidth]{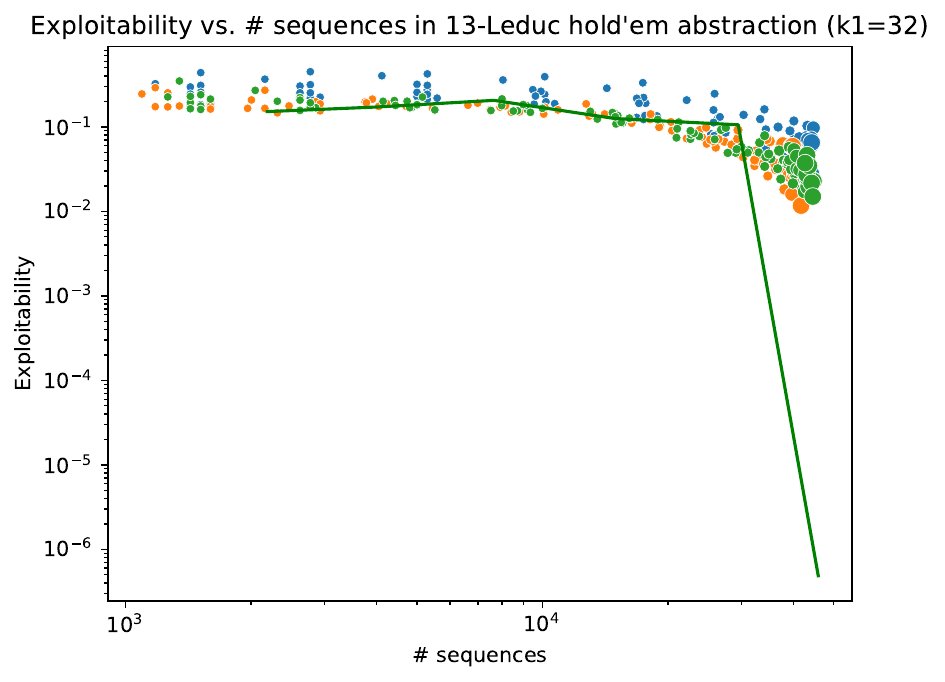}
	\includegraphics[width=0.49\linewidth]{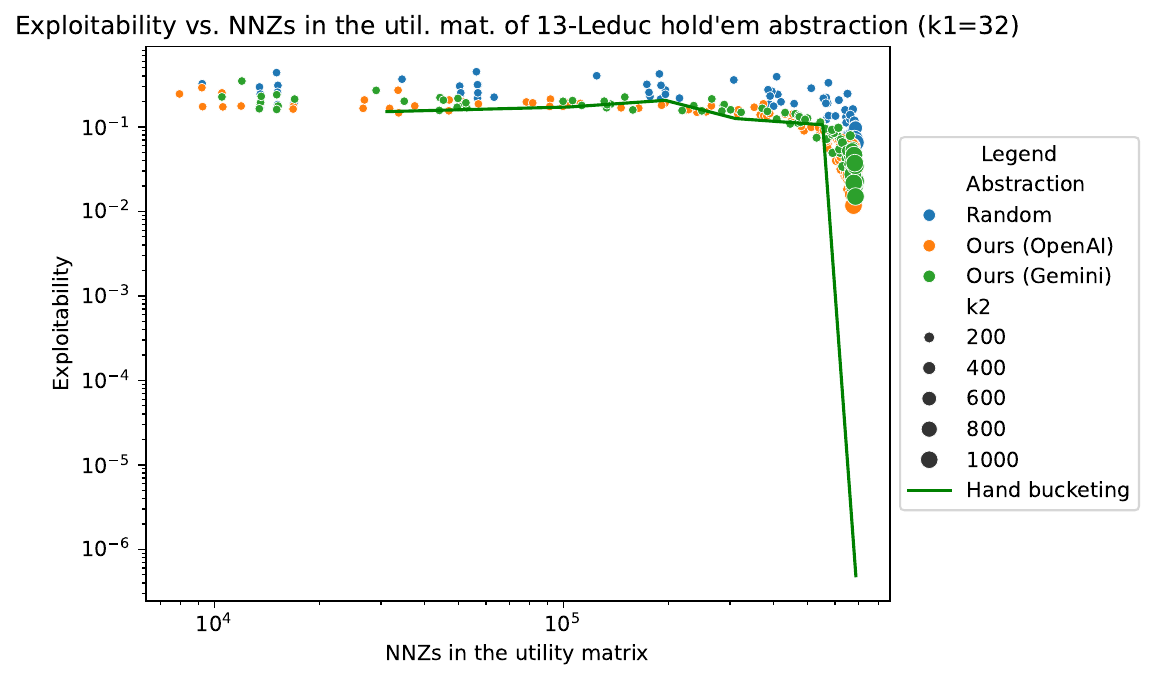}
	\caption{
		Exploitability versus abstraction size in 13-Leduc hold'em.
		Each row shows the results for different $k$ values (for preflop observation clustering).
		Each dot represents one generated abstraction.
		Plots in the same row show the same set of results, but the left one uses the number of sequences for the abstraction size, whereas the right one uses the NNZs in the utility matrix.
	}
	\label{fig:exploitability-leduc}
\end{figure}

Next, we continue by demonstrating the use of foundational embedding models for generating action embeddings. We do this using 13-Leduc poker.
Leduc poker consists of two betting rounds and hence four types of dealing actions to be clustered, one for each combination of players and betting rounds.
We use the same clusters for both players, so the number of types reduces to two -- one for each betting round.
Just as before, we used the poker action embeddings we fetched from OpenAI and Google Gemini from our observational study in Section~\ref{sec:poker-embeddings}.
In generating our abstractions, we ran $k$-means clustering twice, once for each betting round.
For preflop, we used exponentially increasing values of $k_1$, from $1$ to $32$, and, for the flop, we varied $k_2$ to be powers of two from $1$ to $1024$.
Again, for each set of choices for $(k_1, k_2)$, we generated ten separate abstractions with varying random seeds.
We also ran the random baseline using each parameter regime.
The same $(k_1, k_2)$ values were used as the number of buckets for the hand-bucketing baseline.

The exploitability plots for the different abstraction methodologies and choices for $k_1$ on 13-Kuhn poker are shown in Figure~\ref{fig:exploitability}.
It shows how exploitability varies as the size of the game changes for different choices of $k_1$ (for preflop).
For each chosen value for $k_1$, we produced two plots on the same data, as we proposed two different metrics for the size of the abstraction.

As expected, increasing the abstraction size generally decreases exploitability for our technique and the hand-bucketing baseline.
Intuitively, having finer-grained abstraction is beneficial in that it improves the quality of the resulting abstraction.
Unlike in 256-Kuhn poker, increasing the number of clusters slightly helps with the performance of the random baseline as well.
We also note that our technique generally outperforms the random baseline, but the difference does not widen as the resulting abstraction size increases here.
Curiously, the exploitability from action embeddings fetched from Gemini is more tightly concentrated than that using OpenAI embeddings.
This suggests that the use of Gemini models can create abstractions that perform more consistently.

The fact that using foundational embedding models can outperform the random baseline is empirical evidence that it can yield effective abstractions.
That said, we also see that the hand-bucketing baseline, specialized for poker, almost always outperforms our technique.
Again, this is expected since our technique is domain-independent and lacks specialized knowledge about the game structure.
Additional results, including those with other values of $k_1$, are available in Appendix~\ref{sec:additional-leduc}.

\section{Discussion}
\label{sec:attack}

A distinct advantage of generating action embeddings using word embedding techniques is that one can avoid conducting extensive analysis of a game to come up with appropriate metrics or abstraction algorithms.
Indeed, the only requirement of using word vector techniques such as GloVe is the presence of a large dataset of gameplay data, and even this can be ruled out with the use of foundational embedding models, but doing so requires minimal domain-specific knowledge in coming up with common-sense textual representation for the actions, which may not always be possible.

In using our technique, it may be difficult to obtain a sufficient amount of gameplay data, using which word vectors can be fitted.
Also, generating synthetic data using a strategy profile requires one to compute one in the first place, which creates a chicken-and-egg problem (although an approximate strategy profile can possibly be used).
For this, we suggest using human gameplay, and our observational study on chess and experiments on play quality using foundational embedding models suggest that this can result in an effective abstraction.
Moreover, there are situations where fitting action embeddings on human gameplay may be much more desirable than using data generated from an equilibrium strategy profile, namely finding an \textit{agent quantal-response equilibrium (AQRE)}~\cite{mckelveyetal}.
Just like Nash equilibrium, AQRE is a common solution concept for games, whose goal is to capture human behavior and decision-making.
By learning from human gameplay, one can potentially model how humans implicitly perform game abstraction while playing games.

Furthermore, owing to its effective performance, our technique can also serve as a competitive baseline for the development of specialized game abstraction algorithms for specific games.

\section{Conclusions and future research}

We proposed applying word embedding techniques to obtain a vector representation of actions, using which strategically similar actions can be clustered to facilitate game abstraction.
To justify the use of word embedding techniques, we explored the use of word embedding techniques in the context of game playing and demonstrated that action embeddings exhibit highly interpretable results, where actions with similar strategic implications are located closer to each other in the vector space than those that are not.
We also showed that the result remains highly interpretable even when foundational embedding models are used, which are trained on a corpus containing a vast amount of data from multiple domains.
That said, using foundational models requires weakening the domain-independence claim of our contribution, as actions must be converted to a common-sense textual representation.
We contend that this is simpler than developing specialized abstraction algorithms for a particular game.
To demonstrate our technique's effectiveness, we conducted experiments where we calculated the exploitabilities of Nash equilibria for different abstractions when they are lifted to the original game, using GloVe and two leading third-party foundational embedding models.
We also compared our technique with different baselines and found that our technique can produce effective abstractions, but it does not outperform specialized algorithms tailored for specific games. Hence, there is a tradeoff between generality and solution quality.
Another field greatly impacted by the advent of word embedding techniques is the field of information retrieval (IR).
For future work, we suggest applying the ideas from our paper to solve problems in the intersection of computer games and IR.

\section*{Acknowledgements}

This work has been supported by the Vannevar Bush Faculty Fellowship ONR N00014-23-1-2876, National Science Foundation grant RI-2312342, and NIH award A240108S001.
Any opinions, findings, and conclusions or recommendations expressed in this material are those of the authors and do not necessarily reflect the views of the funding agencies.

\bibliographystyle{abbrvnat}
\bibliography{neurips_2026}

@article{brownandsandholm2018,
author = {Noam Brown  and Tuomas Sandholm },
title = {Superhuman {AI} for heads-up no-limit poker: {L}ibratus beats top professionals},
journal = {Science},
volume = {359},
number = {6374},
pages = {418-424},
year = {2018},
}

@article{
brownandsandholm2019,
author = {Noam Brown  and Tuomas Sandholm },
title = {Superhuman {AI} for multiplayer poker},
journal = {Science},
volume = {365},
number = {6456},
pages = {885-890},
year = {2019},
}

@inproceedings{farinaetal2019,
  title = 	 {Regret Circuits: Composability of Regret Minimizers},
  author =       {Farina, Gabriele and Kroer, Christian and Sandholm, Tuomas},
  booktitle = 	 {Proceedings of the International Conference on Machine Learning (ICML)},
  year = 	 {2019}
}

@inproceedings{gilpinandsandholm2006,
author = {Gilpin, Andrew and Sandholm, Tuomas},
title = {A competitive {T}exas {H}old'em poker player via automated abstraction and real-time equilibrium computation},
year = {2006},
booktitle = {Proceedings of the AAAI Conference on Artificial Intelligence (AAAI)},
}

@inproceedings{gilpinandsandholm2007,
author = {Gilpin, Andrew and Sandholm, Tuomas},
title = {Better automated abstraction techniques for imperfect information games, with application to {T}exas {H}old'em poker},
year = {2007},
booktitle = {Proceedings of the International Joint Conference on Autonomous Agents and Multiagent Systems (AAMAS)},
}

@inproceedings{gilpinetal2007,
author = {Gilpin, Andrew and Sandholm, Tuomas and S\o{}rensen, Troels Bjerre},
title = {Potential-aware automated abstraction of sequential games, and holistic equilibrium analysis of {T}exas {H}old'em poker},
year = {2007},
booktitle = {Proceedings of the AAAI Conference on Artificial Intelligence (AAAI)},
}

@inproceedings{gilpinandsandholm2008,
author = {Gilpin, Andrew and Sandholm, Thomas},
title = {Expectation-based versus potential-aware automated abstraction in imperfect information games: an experimental comparison using poker},
year = {2008},
booktitle = {Proceedings of the AAAI Conference on Artificial Intelligence (AAAI)},
}

@article{ganzfriedandsandholm, title={Potential-Aware Imperfect-Recall Abstraction with Earth Mover’s Distance in Imperfect-Information Games}, journal={Proceedings of the AAAI Conference on Artificial Intelligence (AAAI)}, author={Ganzfried, Sam and Sandholm, Tuomas}, year={2014} }

@article{gilpinandsandholm2007b,
author = {Gilpin, Andrew and Sandholm, Tuomas},
title = {Lossless abstraction of imperfect information games},
year = {2007},
volume = {54},
number = {5},
journal = {J. ACM},
pages = {1--30},
}

@misc{neelakantanetal,
      title={Text and Code Embeddings by Contrastive Pre-Training}, 
      author={Arvind Neelakantan and Tao Xu and Raul Puri and Alec Radford and Jesse Michael Han and Jerry Tworek and Qiming Yuan and Nikolas Tezak and Jong Wook Kim and Chris Hallacy and Johannes Heidecke and Pranav Shyam and Boris Power and Tyna Eloundou Nekoul and Girish Sastry and Gretchen Krueger and David Schnurr and Felipe Petroski Such and Kenny Hsu and Madeleine Thompson and Tabarak Khan and Toki Sherbakov and Joanne Jang and Peter Welinder and Lilian Weng},
      year={2022},
}

@inproceedings{penningtonetal,
    title = "{G}lo{V}e: Global Vectors for Word Representation",
    author = "Pennington, Jeffrey  and
      Socher, Richard  and
      Manning, Christopher",
    booktitle = "Proceedings of the Conference on Empirical Methods in Natural Language Processing ({EMNLP})",
    year = "2014",
}

@misc{mikolovetal,
      title={Efficient Estimation of Word Representations in Vector Space}, 
      author={Tomas Mikolov and Kai Chen and Greg Corrado and Jeffrey Dean},
      year={2013},
}

@inproceedings{
	zinkevichetal2007,
	author = {Zinkevich, Martin and Johanson, Michael and Bowling, Michael and Piccione, Carmelo},
	booktitle = {Proceedings of the Annual Conference on Neural Information Processing Systems (NeurIPS)},
	pages = {},
	title = {Regret Minimization in Games with Incomplete Information},
	year = {2007},
}

@misc{pgnmentor,
	author = {{Seqaeon}},
	title = {{PGN} mentor dataset pgns},
	year = {2025},
}

@article{jolliffe,
    author = {Jolliffe, Ian T. and Cadima, Jorge},
    title = {Principal component analysis: a review and recent developments},
    journal = {Philosophical Transactions of the Royal Society A: Mathematical, Physical and Engineering Sciences},
    volume = {374},
    number = {2065},
    year = {2016},
    pages={1--16},
}

@article{kuhn,
    author = {Kuhn, Harold W.},
    title = {Simplified Two-Person Poker},
    journal = {Contributions to the Theory of Games},
    volume = {1},
    year = {1950},
    pages={97--103},
}

@article{mckelveyetal,
author={Mckelvey, Richard D.
and Palfrey, Thomas R.},
title={Quantal Response Equilibria for Extensive Form Games},
journal={Experimental Economics},
year={1998},
volume={1},
number={1},
pages={9--41},
}

@misc{carlsonetal,
      title={A New Pair of {G}lo{V}es}, 
      author={Riley Carlson and John Bauer and Christopher D. Manning},
      year={2025},
}

@inproceedings{waughetal,
author = {Kevin Waugh and David Schnizlein and Michael Bowling and Duane Szafron},
title = {Abstraction Pathologies in Extensive Games},
year = {2009},
booktitle = {Proceedings of the International Conference on Autonomous Agents and Multiagent Systems (AAMAS)},
}

@techreport{johanson,
      title={Measuring the Size of Large No-Limit Poker Games}, 
      author={Michael Johanson},
      year={2013},
      number={TR13-01},
      institution={Department of Computing Science, University of Alberta},
}

@inproceedings{shiandlittman,
author = {Shi, Jiefu and Littman, Michael L.},
title = {Towards Approximately Optimal Poker},
year = {2000},
booktitle = {Proceedings of the AAAI Conference on Artificial Intelligence and Conference on Innovative Applications of Artificial Intelligence (AAAI-IAAI)},
}

@inproceedings{billingsetal,
author = {Billings, D. and Burch, N. and Davidson, A. and Holte, R. and Schaeffer, J. and Schauenberg, T. and Szafron, D.},
title = {Approximating game-theoretic optimal strategies for full-scale poker},
year = {2003},
booktitle = {Proceedings of the International Joint Conference on Artificial Intelligence (IJCAI)},
}

@inproceedings{brownetal,
author = {Brown, Noam and Ganzfried, Sam and Sandholm, Tuomas},
title = {Hierarchical Abstraction, Distributed Equilibrium Computation, and Post-Processing, with Application to a Champion No-Limit {T}exas Hold'em Agent},
year = {2015},
booktitle = {Proceedings of the International Conference on Autonomous Agents and Multiagent Systems (AAMAS)},
}

@inproceedings{sandholm,
author = {Sandholm, Tuomas},
title = {Abstraction for solving large incomplete-information games},
year = {2015},
booktitle = {Proceedings of the AAAI Conference on Artificial Intelligence (AAAI)},
}

@inproceedings{sandholmandsingh,
author = {Sandholm, Tuomas and Singh, Satinder},
title = {Lossy stochastic game abstraction with bounds},
year = {2012},
booktitle = {Proceedings of the ACM Conference on Electronic Commerce (EC)},
}

@inproceedings{kroerandsandholm2014,
author = {Kroer, Christian and Sandholm, Tuomas},
title = {Extensive-form game abstraction with bounds},
year = {2014},
booktitle = {Proceedings of the ACM Conference on Economics and Computation (EC)},
}

@inproceedings{kroerandsandholm2015,
author = {Kroer, Christian and Sandholm, Tuomas},
title = {Discretization of Continuous Action Spaces in Extensive-Form Games},
year = {2015},
booktitle = {Proceedings of the International Conference on Autonomous Agents and Multiagent Systems (AAMAS)},
}

@inproceedings{kroerandsandholm2016,
author = {Kroer, Christian and Sandholm, Tuomas},
title = {Imperfect-Recall Abstractions with Bounds in Games},
year = {2016},
booktitle = {Proceedings of the ACM Conference on Economics and Computation (EC)},
}

@inproceedings{ganzfriedandsandholm2008,
author = {Ganzfried, Sam and Sandholm, Tuomas},
title = {Computing an approximate jam/fold equilibrium for 3-player no-limit {T}exas Hold'em tournaments},
year = {2008},
booktitle = {Proceedings of the International Foundation for Autonomous Agents and Multiagent Systems (AAMAS)},
}

@inproceedings{gilpinetal2008,
author = {Gilpin, Andrew and Sandholm, Tuomas and S\o{}rensen, Troels Bjerre},
title = {A heads-up no-limit {T}exas Hold'em poker player: discretized betting models and automatically generated equilibrium-finding programs},
year = {2008},
booktitle = {Proceedings of the International Foundation for Autonomous Agents and Multiagent Systems (AAMAS)},
}

@inproceedings{southeyetal,
author = {Southey, Finnegan and Bowling, Michael and Larson, Bryce and Piccione, Carmelo and Burch, Neil and Billings, Darse and Rayner, Chris},
title = {{B}ayes' bluff: opponent modelling in poker},
year = {2005},
booktitle = {Proceedings of the Conference on Uncertainty in Artificial Intelligence (UAI)},
}

@InProceedings{shiandlittman2001,
author="Shi, Jiefu and Littman, Michael L.",
title="Abstraction Methods for Game Theoretic Poker",
booktitle="Proceedings of Computers and Games (CG)",
year="2001",
}

@inproceedings{kroerandsandholm2018,
 author = {Kroer, Christian and Sandholm, Tuomas},
 booktitle={Proceedings of the Annual Conference on Neural Information Processing Systems (NeurIPS)},
 title = {A Unified Framework for Extensive-Form Game Abstraction with Bounds},
 year = {2018}
}


\newpage
\appendix

\section{GloVe parameters}
\label{sec:parameters}

In training the GloVe word vectors to generate action embeddings, we used the set of parameters used by~\citet{carlsonetal} for training 50-dimensional word vectors.
The parameters are as follows:
\begin{lstlisting}
VERBOSE=2
MEMORY=4.0
VOCAB_MIN_COUNT=20
VECTOR_SIZE=50
MAX_ITER=100
WINDOW_SIZE=10
BINARY=2
NUM_THREADS=8
X_MAX=10
SEED=42
ALPHA=0.75
ETA=0.075
\end{lstlisting}

We provide an excerpt of the 2.9 GB corpus used to train chess action embeddings below.
\begin{lstlisting}[breaklines,postbreak=\mbox{\textcolor{red}{$\hookrightarrow$}\space}]
c4 g6 e4 Bg7 d4 d6 Nc3 Nf6 Be2 O-O Bg5 Na6 Nf3 h6 Bf4 e5 dxe5 Nh5 Be3 dxe5 Qc1 Kh7 O-O c6 c5 Qe7 Nd2 Nf4 Bxa6 bxa6 Nc4 Qe6 Bxf4 exf4 Nd6 Be5 Rd1 f3 Qe3 fxg2 f4 Bg7 Rd2 Rb8 Re1 Rb4 a3 Rb3 Rxg2 Qe7 e5 Be6 Qg3 Qd7 Qh4 Rfb8 Ree2 Bh3 Rg3 Bf5 Rgg2 Qe6 Nce4 Bxe4 Nxe4 Qd5 h3 Rd3 Kh2 Rbb3 Nc3 Qe6 Rd2 Rxd2 Rxd2 g5 fxg5 Qxe5+ Kg2 Qxg5+ Qxg5 hxg5 Nd1 Be5 Rd7 Kg6 Rxa7 Rg3+ Kf1 Rxh3 Rxa6 Rh1+ Ke2 Rh2+ Kf3 f5 Rxc6+ Kh5 Re6 g4+ Ke3 Bg3 Re8 Rc2 b4 Ra2 c6 Rxa3+ Kd4 Bf4 Nc3 g3 Nd5 Bd6 Rg8 Rb3 c7 Bxc7 Nxc7 Rxb4+ 1/2-1/2
\end{lstlisting}

An excerpt of the corpus of one billion poker hands used to train 256-Kuhn poker embeddings is shown below.
\begin{lstlisting}
57? ?12 c B f -1,1
52? ?44 c C 1,-1
47? ?111 c C -1,1
\end{lstlisting}

\section{Foundational embedding models}
\label{sec:foundational-embedding-models}

We used two third-party foundational embedding models in our paper.
The first was OpenAI's foundational embedding model, named `text-embedding-3-small'.
The second was Google Gemini's premier foundational embedding model, named `gemini-embedding-001', without any specialization for a particular task.

\section{Testbench specification}
\label{sec:resources}

Our testbench has an AMD Ryzen 9 3900X 12-core, 24-thread CPU and 128 GB of memory.

\section{Additional results for $k$-nearest neighbors on chess}
\label{sec:knn}

\begin{table}[h!]
	\caption{Nearest neighbors of chess actions `\texttt{exd8=Q}', `\texttt{Qf7+}', and `\texttt{Bxb5}' from action embeddings obtained from GloVe (left), OpenAI (middle), and Google Gemini (right).}
	\label{tab:knn-all}
	\centering
	\resizebox{0.32\linewidth}{!}{
		\begin{tabular}{c|ccc}
			\toprule
			0. & exd8=Q & Qf7+ & Bxb5 \\
			\midrule
			1. & exd8=Q+ & Qg8+ & Nxb5 \\
			2. & Rcxd8 & Qg6+ & cxb5 \\
			3. & exf8=Q+ & Qe6+ & axb5 \\
			4. & Rbxd8 & Qg7+ & Nb5 \\
			5. & exf8=N+ & Qf8+ & Qxb5 \\
			6. & exf8=R+ & Qf5+ & Bxb5+ \\
			7. & cxd8=Q & Qd7+ & Bb5 \\
			\bottomrule
		\end{tabular}
	}
	\resizebox{0.327\linewidth}{!}{
		\begin{tabular}{c|ccc}
			\toprule
			0. & exd8=Q & Qf7+ & Bxb5 \\
			\midrule
			1. & exd8=Q+ & Qf7 & Bxb5+ \\
			2. & exd8=Q\# & Qf7\# & Bxb5\# \\
			3. & exf8=Q & Qxf7+ & Bxb6 \\
			4. & exd8=R+ & Qxf7 & Bxb6+ \\
			5. & exf8=Q\# & Qf5+ & Bxb3 \\
			6. & exf8=Q+ & Qh7+ & Bxb4 \\
			7. & dxe8=Q & Qf5 & Bxb3+ \\
			\bottomrule
		\end{tabular}
	}
	\resizebox{0.327\linewidth}{!}{
		\begin{tabular}{c|ccc}
			\toprule
			0. & exd8=Q & Qf7+ & Bxb5 \\
			\midrule
			1. & exd8=Q\# & Qgf7+ & Bxb5\# \\
			2. & exd8=Q+ & Qhf7+ & Bxb5+ \\
			3. & exd1=Q &  Qf7\# & Bxb2 \\
			4. & exd1=Q\# & Qf7 & Bxb6 \\
			5. & exd8=R+ & Qf3+ & Rbxb5 \\
			6. & exf8=Q &  Qfg7+ & Bxb1 \\
			7. & exd8=N+ & Qf6+ & Bxb8 \\
			\bottomrule
		\end{tabular}
	}
\end{table}

The nearest neighbors of the target chess actions we chose using embeddings from GloVe, OpenAI `\texttt{text-embedding-3-small}' and Google Gemini `\texttt{gemini-embedding-001}' are shown in Table~\ref{tab:knn-all}.
Recall that we obtained the nearest neighbors only for the vocabulary from the GloVe word vectors.
Thus, there may exist other chess actions that are closer than some of the neighbors mentioned here.

\section{Benchmark games}
\label{sec:games}

In this section, we define a couple of games that we used to test our technique.

\subsection{256-Kuhn poker}

256-Kuhn poker is a variant of Kuhn poker~\cite{kuhn} that increases the number of cards in the deck to 256 (from just a Jack, Queen, and King).
The play begins by both players committing a single chip to the pot.
The dealer then randomly samples two cards (without replacement) from the deck and gives one to each player as their hole card.
Each player can only observe their own hole card.

A single betting round ensues, where players can choose among the following actions: check, call, bet, and fold.
To check, no player must have previously made a bet.
It signals that the player does not commit any more money into the pot.
When calling (which is only possible when there is an outstanding bet by a player), the player must match the bet made by the other player.
When betting, the player increases the stakes by placing a bet of a single chip, forcing the other player to either match it or abandon the pot.
A player is said to fold when they abandon the pot, unable to match an outstanding bet of their opponent.
Kuhn poker only allows one bet in the betting round.

When both players are in the pot even after the betting ends, a showdown ensues, where both players reveal their private cards to compare them.
The victor is the player with a higher card, who then takes the money committed by both players.

\subsection{13-Leduc hold'em}

13-Leduc hold'em is a variant of Leduc hold'em~\cite{southeyetal} that increases the range of ranks in the deck from just Jack, Queen, and King to deuces to Aces.
There are two cards in the deck for every rank, with hearts and spades as suits.
Just as in 256-Kuhn poker, the play begins by both players committing a single chip to the pot.
The dealer then randomly samples two cards (without replacement) from the deck and gives one to each player as their hole card.
Again, each player can observe their own card, but is not aware of the card owned by the other.

A betting round ensues, which is identical to that of 256-Kuhn poker but with some differences.
First, two bets are allowed, as opposed to just one, and the bet size is 2 chips.
This phase of the game is denoted as the preflop.
Then, the dealer reveals a single public card among the remaining set of cards in the deck, after which another betting round ensues.
Again, two bets are allowed, and the bet size used in this phase is 4 chips.
This phase of the game is denoted as the flop.

As before, showdown ensues when both players remain in the pot after the two betting rounds.
A player who has paired their card with the public card wins the pot.
If no player paired their hole card, the player with the higher hole card wins.

\section{Additional results for $k$-nearest neighbors on 256-Kuhn poker}
\label{sec:knn-kuhn}

\begin{table}[h!]
	\centering
	\caption{
		Nearest neighbors of dealing the row player 0\textsuperscript{th}, 127\textsuperscript{th}, and 257\textsuperscript{th} cards from action embeddings obtained using GloVe.
		The neighbors are sorted from closest to furthest.
		Each number represents the ordinal of the corresponding card of a neighbor.
	}
	\label{tab:knn-kuhn}
	\begin{tabular}{c|ccc}
		\toprule
		0. & 0\textsuperscript{th} & 127\textsuperscript{th} & 255\textsuperscript{th} \\
		\midrule
		1. & 15\textsuperscript{th} & 126\textsuperscript{th} & 23\textsuperscript{rd} \\
		2. & 22\textsuperscript{nd} & 125\textsuperscript{th} & 220\textsuperscript{th} \\
		3. & 6\textsuperscript{th} & 124\textsuperscript{th} & 222\textsuperscript{nd} \\
		4. & 16\textsuperscript{th} & 122\textsuperscript{nd} & 224\textsuperscript{th} \\
		5. & 9\textsuperscript{th} & 123\textsuperscript{rd} & 229\textsuperscript{th} \\
		6. & 17\textsuperscript{th} & 121\textsuperscript{st} & 221\textsuperscript{st} \\
		7. & 12\textsuperscript{th} & 119\textsuperscript{th} & 218\textsuperscript{th} \\
		\bottomrule
	\end{tabular}
\end{table}

The nearest neighbors of the target dealing actions in 256-Kuhn poker we chose using GloVe are shown in Table~\ref{tab:knn-kuhn}.
Interpretation of this data, along with the visualization, is detailed in Section~\ref{sec:observational-studies}.
 
\section{Additional results for $k$-nearest neighbors on Texas hold'em}
\label{sec:knn-leduc}

\begin{table}[h!]
	\caption{Nearest neighbors of dealing the row player `AsAh', `JsTs', and `2c7d' from action embeddings obtained from OpenAI (left) and Google Gemini (right).}
	\label{tab:knn-leduc}
	\centering
	\begin{tabular}{c|ccc}
		\toprule
		0. & AsAh & JsTs & 2c7d \\
		\midrule
		1. & AhAs & JsTc & 2c7c \\
		2. & AhKh & ThJs & 4c7d \\
		3. & AcAh & JhTs & 3c7d \\
		4. & ThAh & JsTh & 2c6d \\
		5. & QsAh & JsTd & 5c7d \\
		6. & QhAh & TsJs & 2c7h \\
		7. & AsKh & JdTs & 7c2c \\
		\bottomrule
	\end{tabular}
	\hspace{1em}
	\begin{tabular}{c|ccc}
		\toprule
		0. & AsAh & JsTs & 2c7d \\
		\midrule
		1. & AdAh & JcTs & 3c7d \\
		2. & AsJh & JdTs & 7c7d \\
		3. & JsAh & JhTs & 6c7d \\
		4. & As3h & KsTs & 4c7d \\
		5. & AsKh & JsTh & 2c7h \\
		6. & KsAh & 5sTs & 2c7s \\
		7. & AcAh & JsTc & 5c7d \\
		\bottomrule
	\end{tabular}
\end{table}

The nearest neighbors of the target dealing actions in Texas hold'em we chose using foundational embedding models from OpenAI and Google Gemini are shown in Table~\ref{tab:knn-leduc}.
Interpretation of this data, along with the visualization, is detailed in Section~\ref{sec:observational-studies}.

\section{Textual representation of poker hands}
\label{sec:poker-hands}

Poker hands consist of poker cards, each of which, in turn, consists of a rank and a suit.
The possible ranks are deuces, treys, fours, and so on until tens, then Jacks, Queens, Kings, and Aces.
These ranks are represented using characters from `2' to `9', then `T', `J', `Q', `K', and `A', respectively.
The possible suits are clubs, diamonds, hearts, and spades, which are usually represented using characters `c', `d', `h', and `s', respectively.
To obtain a textual representation of a single poker card, the characters for its rank and suit are concatenated (\textit{e.g.}, `As' for the Ace of spades).
A poker hand can then be textually represented by concatenating the texts of its component cards.

\section{Additional experimental results on play quality}
\label{sec:additional-leduc}

\begin{figure}[h!]
	\centering
	\includegraphics[width=0.4\linewidth]{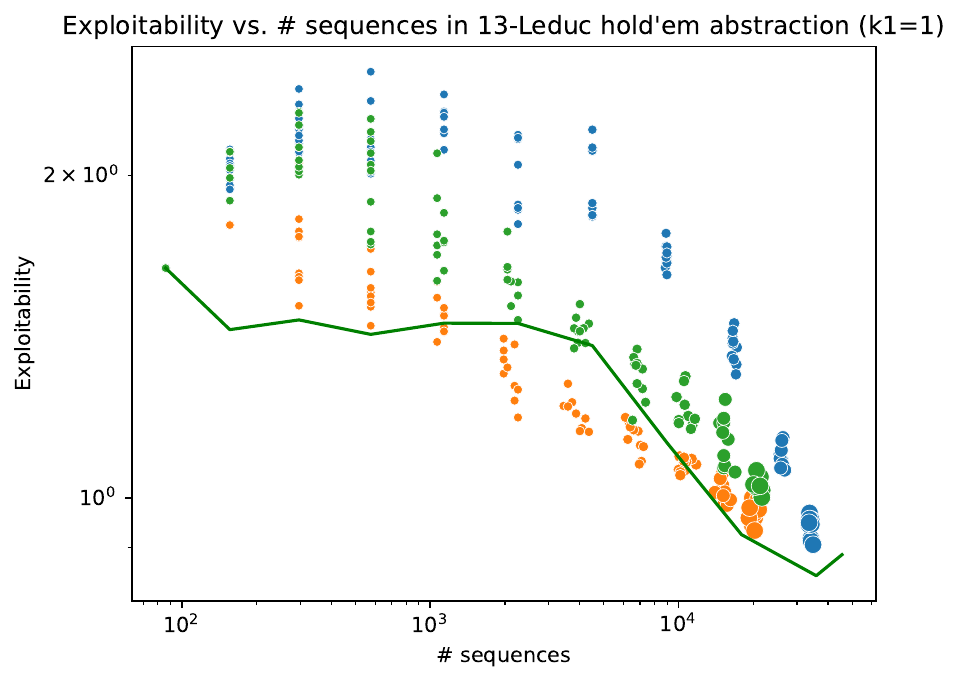}
	\includegraphics[width=0.478\linewidth]{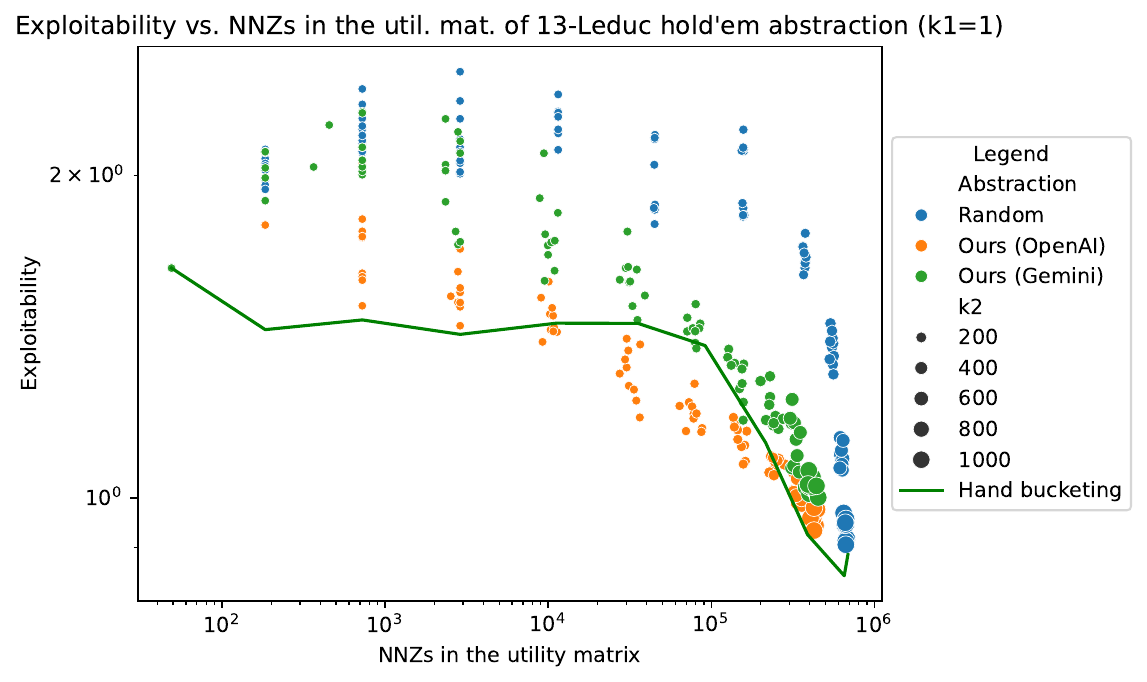}
	\includegraphics[width=0.4\linewidth]{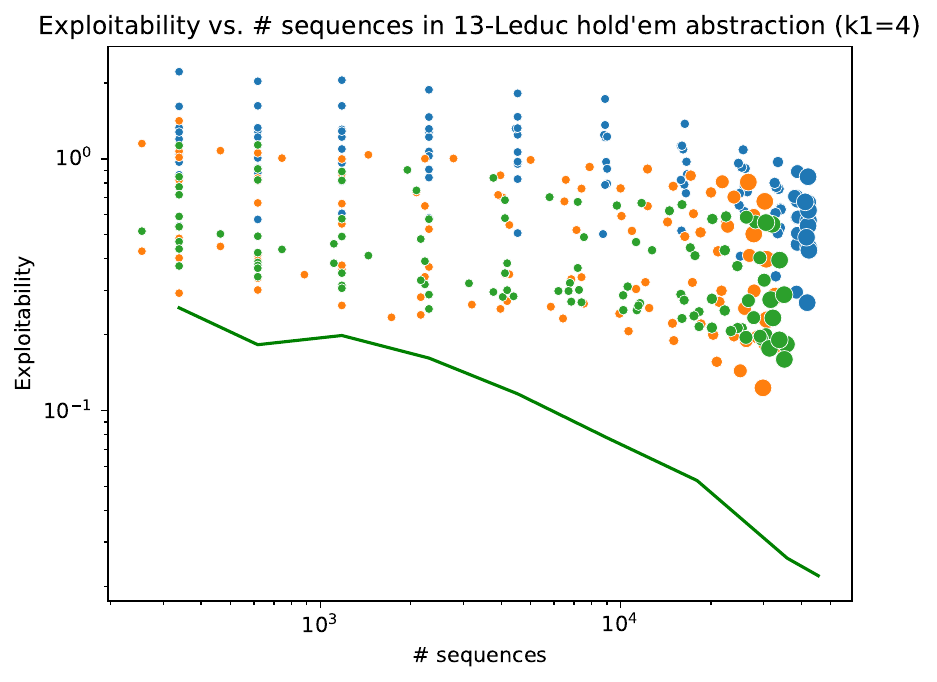}
	\includegraphics[width=0.49\linewidth]{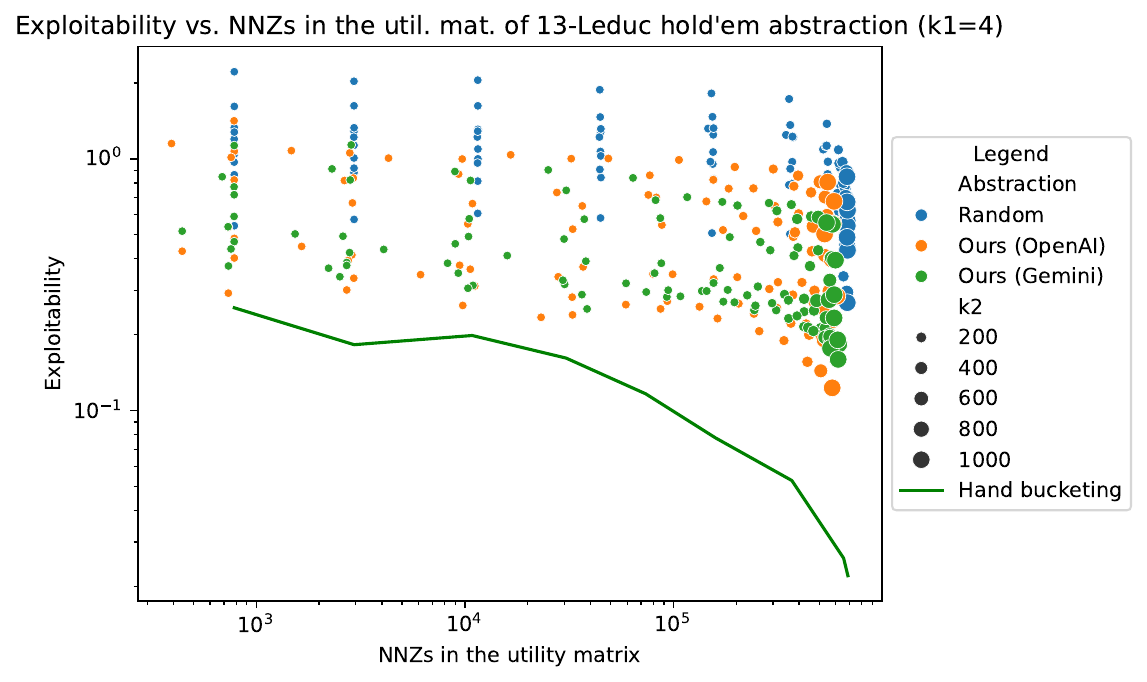}
	\includegraphics[width=0.4\linewidth]{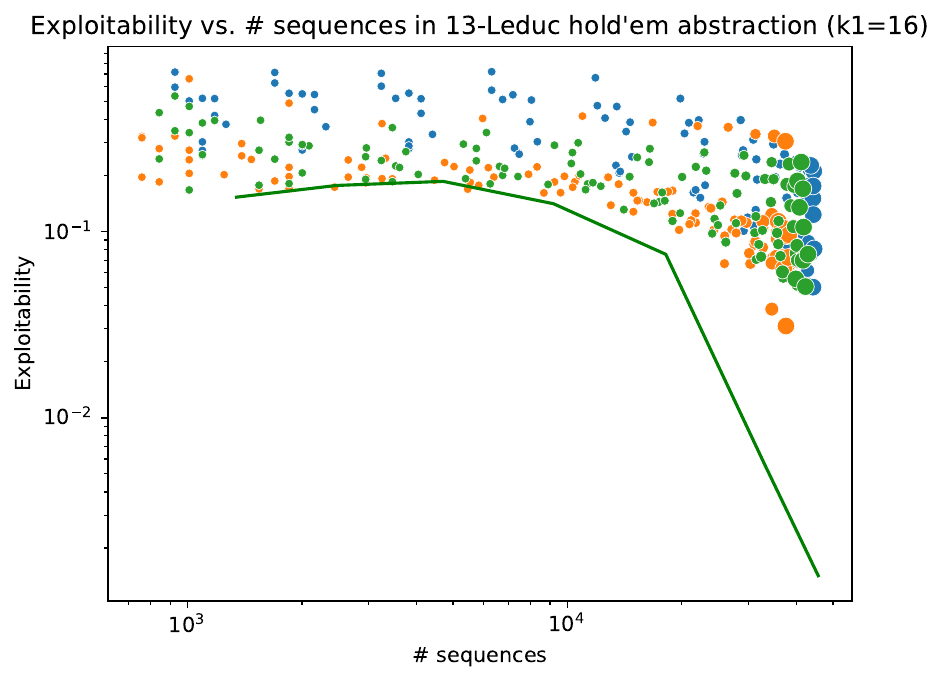}
	\includegraphics[width=0.49\linewidth]{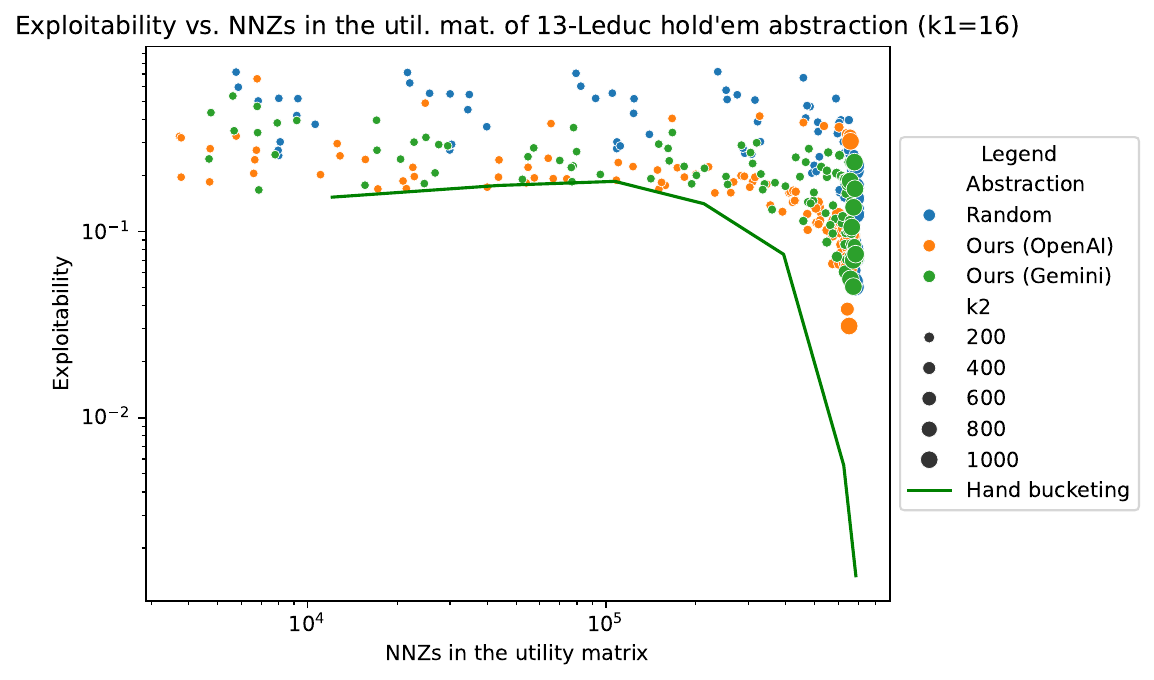}
	\caption{
		Exploitability versus abstraction size in 13-Leduc hold'em.
		Each row shows the results for different $k$ values (for preflop observation clustering).
		Each dot represents one generated abstraction.
		Plots in the same row show the same set of results, but the left one uses the number of sequences for the abstraction size, whereas the right one uses the NNZs in the utility matrix.
	}
	\label{fig:exploitability-leduc2}
\end{figure}

The exploitability plots for the different abstraction methodologies and other choices for $k_1$ on 13-Kuhn poker are shown in Figure~\ref{fig:exploitability}.
Our technique is observed to outperform the random baseline, but it does worse than the hand-bucketing baseline, which is tailored for poker.
We noticed that the larger choices for $k$ for $k$-means clustering resulted in empty clusters due to the algorithm's inherently random nature.
In fact, for 256-Kuhn poker, we observed this for choices of $k$ as low as even $k = 4$ for one of the seeds.
Thus, increasing $k$ after a certain threshold seldom helped create finer-grained abstraction; this is not an issue, as empty clusters do not contribute to the abstraction size.

For each game abstraction algorithm, we also tabulated the mean exploitabilities using our technique and baselines, along with their standard errors of the mean, as shown in Tables~\ref{tab:kuhn-exps},~\ref{tab:exploitability-leduc-random},~\ref{tab:exploitability-leduc-openai},~\ref{tab:exploitability-leduc-google}, and~\ref{tab:exploitability-leduc-hand}.
It is \textbf{important} to note that one cannot directly compare the values between the corresponding cells in different tables because the input parameters to the abstraction algorithms do \textbf{not} reflect the size of the resulting abstraction.
Thus, to properly compare these techniques, one must consult their exploitability-size plots, shown in Figures~\ref{fig:exploitability},~\ref{fig:exploitability-leduc}, and~\ref{fig:exploitability-leduc2}.

\newpage

\begin{table}[H]
	\caption{
		Exploitabilities of lifted Nash equilibria in 256-Kuhn poker from abstractions produced by the random baseline (left), our technique (middle), and the hand-bucketing baseline (right).
		Each value is given with the standard error of the mean.
		The hand-bucketing baseline is deterministic, so the standard error of the mean is not given for the right table.
		Directly comparing values between tables can be misleading.
		To properly compare these techniques, one must consult their exploitability-size plots, shown in Figures~\ref{fig:exploitability-leduc} and~\ref{fig:exploitability-leduc2}.
	}
	\label{tab:kuhn-exps}
	\centering
	\begin{tabular}{rc}
		\toprule
		k & Exploitabilities \\
		\midrule
		1 & $1.887 \times 10^{-1}$ \\
		2 & $0.24 \pm 0.03$ \\
		4 & $0.37 \pm 0.01$ \\
		8 & $0.38 \pm 0.02$ \\
		16 & $0.323 \pm 0.009$ \\
		32 & $0.323 \pm 0.009$ \\
		64 & $0.32 \pm 0.01$ \\
		128 & $0.254 \pm 0.009$ \\
		256 & $0.181 \pm 0.004$ \\
		512 & $0.113 \pm 0.005$ \\
		\bottomrule
	\end{tabular}
	\begin{tabular}{rc}
		\toprule
		k & Exploitabilities \\
		\midrule
		1 & $1.887 \times 10^{-1}$ \\
		2 & $0.143 \pm 0.007$ \\
		4 & $0.29 \pm 0.03$ \\
		8 & $0.27 \pm 0.01$ \\
		16 & $0.18 \pm 0.02$ \\
		32 & $0.06 \pm 0.02$ \\
		64 & $0.035 \pm 0.007$ \\
		128 & $0.010 \pm 0.003$ \\
		256 & $0.009 \pm 0.003$ \\
		512 & $0.007 \pm 0.003$ \\
		\bottomrule
	\end{tabular}
	\resizebox{0.254\linewidth}{!}{
		\begin{tabular}{rc}
			\toprule
			k & Exploitabilities \\
			\midrule
			1 & $1.887 \times 10^{-1}$ \\
			2 & $1.091 \times 10^{-1}$ \\
			4 & $1.712 \times 10^{-2}$ \\
			8 & $7.615 \times 10^{-3}$ \\
			16 & $1.900 \times 10^{-3}$ \\
			32 & $7.510 \times 10^{-4}$ \\
			64 & $1.991 \times 10^{-4}$ \\
			128 & $7.181 \times 10^{-5}$ \\
			256 & $2.949 \times 10^{-16}$ \\
			512 & $2.949 \times 10^{-16}$ \\
			\bottomrule
		\end{tabular}
	}
\end{table}

\begin{table}[H]
	\caption{
		Exploitabilities of lifted Nash equilibria in 13-Leduc poker from abstractions produced by our technique using the random baseline.
		Each value is given with the standard error of the mean.
		Directly comparing values between tables can be misleading.
		To properly compare these techniques, one must consult their exploitability-size plots, shown in Figures~\ref{fig:exploitability-leduc} and~\ref{fig:exploitability-leduc2}.
	}
	\label{tab:exploitability-leduc-random}
	\centering
	\resizebox{\linewidth}{!}{
		\begin{tabular}{c|cccccc}
		\toprule
		& $k_1=1$ & $k_1=2$ & $k_1=4$ & $k_1=8$ & $k_1=16$ & $k_1=32$ \\
		\midrule
		$k_2=1$ & $1.640$ & $1.5 \pm 0.1$ & $1.2 \pm 0.1$ & $0.82 \pm 0.09$ & $0.45 \pm 0.05$ & $0.26 \pm 0.03$ \\
		$k_2=2$ & $2.01 \pm 0.02$ & $1.5 \pm 0.1$ & $1.2 \pm 0.1$ & $0.82 \pm 0.09$ & $0.46 \pm 0.05$ & $0.29 \pm 0.02$ \\
		$k_2=4$ & $2.21 \pm 0.03$ & $1.5 \pm 0.1$ & $1.2 \pm 0.1$ & $0.82 \pm 0.09$ & $0.45 \pm 0.05$ & $0.29 \pm 0.02$ \\
		$k_2=8$ & $2.17 \pm 0.05$ & $1.5 \pm 0.1$ & $1.2 \pm 0.1$ & $0.80 \pm 0.09$ & $0.43 \pm 0.05$ & $0.25 \pm 0.02$ \\
		$k_2=16$ & $2.24 \pm 0.02$ & $1.5 \pm 0.1$ & $1.1 \pm 0.1$ & $0.74 \pm 0.08$ & $0.36 \pm 0.05$ & $0.20 \pm 0.02$ \\
		$k_2=32$ & $2.00 \pm 0.05$ & $1.4 \pm 0.1$ & $1.1 \pm 0.1$ & $0.65 \pm 0.08$ & $0.29 \pm 0.04$ & $0.13 \pm 0.02$ \\
		$k_2=64$ & $1.99 \pm 0.05$ & $1.3 \pm 0.1$ & $0.94 \pm 0.08$ & $0.54 \pm 0.07$ & $0.21 \pm 0.03$ & $0.09 \pm 0.01$ \\
		$k_2=128$ & $1.68 \pm 0.01$ & $1.13 \pm 0.07$ & $0.78 \pm 0.06$ & $0.44 \pm 0.06$ & $0.17 \pm 0.02$ & $0.06 \pm 0.01$ \\
		$k_2=256$ & $1.38 \pm 0.01$ & $0.90 \pm 0.05$ & $0.66 \pm 0.06$ & $0.39 \pm 0.05$ & $0.14 \pm 0.02$ & $0.051 \pm 0.009$ \\
		$k_2=512$ & $1.095 \pm 0.008$ & $0.75 \pm 0.04$ & $0.59 \pm 0.05$ & $0.37 \pm 0.05$ & $0.13 \pm 0.02$ & $0.046 \pm 0.009$ \\
		$k_2=1024$ & $0.941 \pm 0.007$ & $0.69 \pm 0.04$ & $0.56 \pm 0.05$ & $0.35 \pm 0.05$ & $0.13 \pm 0.02$ & $0.043 \pm 0.008$ \\
		\bottomrule
		\end{tabular}
	}
\end{table}

\begin{table}[H]
	\caption{
		Exploitabilities of lifted Nash equilibria in 13-Leduc poker from abstractions produced by our technique using OpenAI embeddings.
		Each value is given with the standard error of the mean.
		Directly comparing values between tables can be misleading.
		To properly compare these techniques, one must consult their exploitability-size plots, shown in Figures~\ref{fig:exploitability-leduc} and~\ref{fig:exploitability-leduc2}.
	}
	\label{tab:exploitability-leduc-openai}
	\centering
	\resizebox{\linewidth}{!}{
		\begin{tabular}{c|cccccc}
		\toprule
		& $k_1=1$ & $k_1=2$ & $k_1=4$ & $k_1=8$ & $k_1=16$ & $k_1=32$ \\
		\midrule
		$k_2=1$ & $1.640$ & $1.2 \pm 0.1$ & $0.8 \pm 0.1$ & $0.5 \pm 0.1$ & $0.30 \pm 0.04$ & $0.20 \pm 0.01$ \\
		$k_2=2$ & $1.799$ & $0.9 \pm 0.1$ & $0.63 \pm 0.09$ & $0.40 \pm 0.07$ & $0.24 \pm 0.03$ & $0.19 \pm 0.01$ \\
		$k_2=4$ & $1.76 \pm 0.06$ & $0.9 \pm 0.1$ & $0.57 \pm 0.09$ & $0.36 \pm 0.06$ & $0.22 \pm 0.02$ & $0.185 \pm 0.005$ \\
		$k_2=8$ & $1.55 \pm 0.02$ & $0.9 \pm 0.1$ & $0.5 \pm 0.1$ & $0.35 \pm 0.06$ & $0.22 \pm 0.02$ & $0.165 \pm 0.005$ \\
		$k_2=16$ & $1.47 \pm 0.02$ & $0.8 \pm 0.1$ & $0.53 \pm 0.09$ & $0.34 \pm 0.06$ & $0.21 \pm 0.02$ & $0.139 \pm 0.007$ \\
		$k_2=32$ & $1.31 \pm 0.02$ & $0.72 \pm 0.09$ & $0.52 \pm 0.09$ & $0.34 \pm 0.06$ & $0.18 \pm 0.02$ & $0.098 \pm 0.008$ \\
		$k_2=64$ & $1.20 \pm 0.01$ & $0.68 \pm 0.09$ & $0.48 \pm 0.08$ & $0.30 \pm 0.06$ & $0.15 \pm 0.02$ & $0.068 \pm 0.006$ \\
		$k_2=128$ & $1.14 \pm 0.01$ & $0.63 \pm 0.08$ & $0.46 \pm 0.08$ & $0.28 \pm 0.06$ & $0.13 \pm 0.03$ & $0.050 \pm 0.006$ \\
		$k_2=256$ & $1.075 \pm 0.005$ & $0.59 \pm 0.07$ & $0.42 \pm 0.08$ & $0.25 \pm 0.05$ & $0.12 \pm 0.02$ & $0.040 \pm 0.005$ \\
		$k_2=512$ & $1.015 \pm 0.007$ & $0.55 \pm 0.07$ & $0.41 \pm 0.07$ & $0.24 \pm 0.05$ & $0.11 \pm 0.03$ & $0.035 \pm 0.005$ \\
		$k_2=1024$ & $0.979 \pm 0.007$ & $0.53 \pm 0.07$ & $0.39 \pm 0.07$ & $0.23 \pm 0.05$ & $0.10 \pm 0.02$ & $0.034 \pm 0.005$ \\
		\bottomrule
		\end{tabular}
	}
\end{table}

\begin{table}[H]
	\caption{
		Exploitabilities of lifted Nash equilibria in 13-Leduc poker from abstractions produced by our technique using Google Gemini embeddings.
		Each value is given with the standard error of the mean.
		Directly comparing values between tables can be misleading.
		To properly compare these techniques, one must consult their exploitability-size plots, shown in Figures~\ref{fig:exploitability-leduc} and~\ref{fig:exploitability-leduc2}.
	}
	\label{tab:exploitability-leduc-google}
	\centering
	\resizebox{\linewidth}{!}{
		\begin{tabular}{c|cccccc}
		\toprule
		& $k_1=1$ & $k_1=2$ & $k_1=4$ & $k_1=8$ & $k_1=16$ & $k_1=32$ \\
		\midrule
		$k_2=1$ & $1.640$ & $1.47 \pm 0.07$ & $0.64 \pm 0.07$ & $0.39 \pm 0.03$ & $0.36 \pm 0.04$ & $0.21 \pm 0.02$ \\
		$k_2=2$ & $2.06 \pm 0.02$ & $1.6 \pm 0.1$ & $0.58 \pm 0.09$ & $0.33 \pm 0.03$ & $0.27 \pm 0.02$ & $0.20 \pm 0.01$ \\
		$k_2=4$ & $2.11 \pm 0.04$ & $1.6 \pm 0.2$ & $0.50 \pm 0.06$ & $0.30 \pm 0.01$ & $0.24 \pm 0.02$ & $0.188 \pm 0.006$ \\
		$k_2=8$ & $1.99 \pm 0.06$ & $1.5 \pm 0.1$ & $0.47 \pm 0.07$ & $0.28 \pm 0.01$ & $0.24 \pm 0.02$ & $0.170 \pm 0.006$ \\
		$k_2=16$ & $1.77 \pm 0.05$ & $1.3 \pm 0.1$ & $0.43 \pm 0.06$ & $0.25 \pm 0.01$ & $0.22 \pm 0.02$ & $0.129 \pm 0.005$ \\
		$k_2=32$ & $1.60 \pm 0.03$ & $1.2 \pm 0.1$ & $0.40 \pm 0.05$ & $0.23 \pm 0.01$ & $0.18 \pm 0.02$ & $0.088 \pm 0.006$ \\
		$k_2=64$ & $1.44 \pm 0.01$ & $1.09 \pm 0.09$ & $0.38 \pm 0.05$ & $0.21 \pm 0.01$ & $0.16 \pm 0.02$ & $0.061 \pm 0.006$ \\
		$k_2=128$ & $1.30 \pm 0.02$ & $0.99 \pm 0.08$ & $0.36 \pm 0.05$ & $0.19 \pm 0.01$ & $0.14 \pm 0.02$ & $0.045 \pm 0.005$ \\
		$k_2=256$ & $1.21 \pm 0.02$ & $0.94 \pm 0.06$ & $0.33 \pm 0.05$ & $0.17 \pm 0.01$ & $0.13 \pm 0.02$ & $0.036 \pm 0.004$ \\
		$k_2=512$ & $1.14 \pm 0.02$ & $0.87 \pm 0.06$ & $0.32 \pm 0.05$ & $0.16 \pm 0.01$ & $0.12 \pm 0.02$ & $0.032 \pm 0.004$ \\
		$k_2=1024$ & $1.027 \pm 0.005$ & $0.80 \pm 0.05$ & $0.30 \pm 0.05$ & $0.16 \pm 0.01$ & $0.12 \pm 0.02$ & $0.027 \pm 0.003$ \\
		\bottomrule
		\end{tabular}
	}
\end{table}

\begin{table}[H]
	\caption{
		Exploitabilities of lifted Nash equilibria in 13-Leduc poker from abstractions produced by the hand-bucketing baseline.
		Directly comparing values between tables can be misleading.
		To properly compare these techniques, one must consult their exploitability-size plots, shown in Figures~\ref{fig:exploitability-leduc} and~\ref{fig:exploitability-leduc2}.
	}
	\label{tab:exploitability-leduc-hand}
	\centering
	\resizebox{\linewidth}{!}{
		\begin{tabular}{c|cccccc}
		\toprule
		& $k_1=1$ & $k_1=2$ & $k_1=4$ & $k_1=8$ & $k_1=16$ & $k_1=32$ \\
		\midrule
		$k_2=1$ & $1.640$ & $6.926 \times 10^{-1}$ & $2.556 \times 10^{-1}$ & $1.765 \times 10^{-1}$ & $1.527 \times 10^{-1}$ & $1.517 \times 10^{-1}$ \\
		$k_2=2$ & $1.437$ & $5.984 \times 10^{-1}$ & $1.824 \times 10^{-1}$ & $1.996 \times 10^{-1}$ & $1.764 \times 10^{-1}$ & $1.720 \times 10^{-1}$ \\
		$k_2=4$ & $1.467$ & $4.881 \times 10^{-1}$ & $1.984 \times 10^{-1}$ & $1.967 \times 10^{-1}$ & $1.856 \times 10^{-1}$ & $2.062 \times 10^{-1}$ \\
		$k_2=8$ & $1.422$ & $3.615 \times 10^{-1}$ & $1.614 \times 10^{-1}$ & $1.582 \times 10^{-1}$ & $1.408 \times 10^{-1}$ & $1.256 \times 10^{-1}$ \\
		$k_2=16$ & $1.457$ & $3.269 \times 10^{-1}$ & $1.164 \times 10^{-1}$ & $8.811 \times 10^{-2}$ & $7.532 \times 10^{-2}$ & $1.059 \times 10^{-1}$ \\
		$k_2=32$ & $1.456$ & $3.204 \times 10^{-1}$ & $7.780 \times 10^{-2}$ & $3.716 \times 10^{-2}$ & $5.554 \times 10^{-3}$ & $4.879 \times 10^{-7}$ \\
		$k_2=64$ & $1.388$ & $2.854 \times 10^{-1}$ & $5.264 \times 10^{-2}$ & $1.942 \times 10^{-2}$ & $1.420 \times 10^{-3}$ & $4.879 \times 10^{-7}$ \\
		$k_2=128$ & $1.127$ & $2.467 \times 10^{-1}$ & $2.587 \times 10^{-2}$ & $1.924 \times 10^{-2}$ & $1.420 \times 10^{-3}$ & $4.879 \times 10^{-7}$ \\
		$k_2=256$ & $9.249 \times 10^{-1}$ & $1.702 \times 10^{-1}$ & $2.204 \times 10^{-2}$ & $1.924 \times 10^{-2}$ & $1.420 \times 10^{-3}$ & $4.879 \times 10^{-7}$ \\
		$k_2=512$ & $8.468 \times 10^{-1}$ & $2.012 \times 10^{-1}$ & $2.204 \times 10^{-2}$ & $1.924 \times 10^{-2}$ & $1.420 \times 10^{-3}$ & $4.879 \times 10^{-7}$ \\
		$k_2=1024$ & $8.859 \times 10^{-1}$ & $2.012 \times 10^{-1}$ & $2.204 \times 10^{-2}$ & $1.924 \times 10^{-2}$ & $1.420 \times 10^{-3}$ & $4.879 \times 10^{-7}$ \\
		\bottomrule
		\end{tabular}
	}
\end{table}



\end{document}